\documentclass{aastex61} 
\usepackage{graphicx}
\begin{document}
\title{The Connection Between Different Tracers \\ Of The Diffuse Interstellar Medium: Kinematics}
\author{Johnathan S. Rice} 
\author{S. R. Federman} 
\affil{University of Toledo Physics and Astronomy, Toledo, OH 43606, USA; Johnathan.Rice@utoledo.edu, Steven.Federman@utoledo.edu 
}

\author{Nicolas Flagey}
\affil{Canada-France-Hawaii Telescope Corporation, 65-1238 Mamalahoa Hwy, Kamuela, Hawaii 96743, USA; Flagey@cfht.hawaii.edu
}

\author{Paul F. Goldsmith}
\author{William D. Langer}
\author{Jorge L. Pineda}
\affil{Jet Propulsion Laboratory, California Institute of Technology, Pasadena, CA 91109, USA; Paul.F.Goldsmith@jpl.nasa.gov, William.D.Langer@jpl.nasa.gov, Jorge.Pineda@jpl.nasa.gov
}

\author{D. L. Lambert}
\affil{W. J. McDonald Observatory and Department of Astronomy, University of Texas at Austin, Austin, TX 78712, USA; DLL@astro.as.utexas.edu 
}

\begin{abstract} 

Using visible, radio, microwave, and sub-mm data, we study several lines of sight toward stars generally closer than 1 kpc on a component-by-component basis. We derive the component structure seen in absorption at visible wavelengths from \ion{Ca}{2}, \ion{Ca}{1}, \ion{K}{1}, CH, CH$^{+}\!,$ and CN and compare it to emission from \ion{H}{1}, CO and its isotopologues, and C$^{+}$  from the GOT C+ survey. The correspondence between components in emission and absorption help create a more unified picture of diffuse atomic and molecular gas in the interstellar medium. We also discuss how these tracers are related to the CO-dark H$_{2}$ gas  probed by C$^{+}$ emission and discuss the kinematic connections among the species observed.
\end{abstract}

\section{Introduction}
The formation of molecular clouds proceeds from  diffuse atomic gas to dense molecular clouds in which star formation takes place. In the earliest stages, the atomic hydrogen in diffuse atomic gas transitions to molecular hydrogen seen in diffuse molecular gas, which contains a mix of atomic and molecular gas. While atomic hydrogen can be observed at 21 cm in emission in this material, molecular hydrogen can only be observed in absorption against a background source. This first transition illustrates that combining the information from different species can trace a larger range of physical conditions and when considered as an ensemble they provide a more complete picture of an entire interstellar cloud. Many times, species observed in absorption are only combined with other species in absorption while species observed in emission are combined with other species in emission, creating two separate views of the same evolution.  We combine the emission and absorption data toward seventeen background sources and the ten nearest GOT C+ lines of sight to begin bridging the gap between two different views of the diffuse interstellar medium (ISM): the paradigm created by tracers seen in absorption at visible wavelengths and the one created by tracers seen in emission at longer wavelengths.

Studies connecting UV/visible absorption and long-wavelength emission began more 40 years ago.  Knapp and Jura (1976) sought CO emission from gas observed in absorption with the $Copernicus$ satellite.  Liszt (1979) acquired data on CH and CO ($J = 1\rightarrow0$) emission along the line of sight to $\zeta$ Oph, finding material with a very small Doppler parameter.  Subsequent observations of CO $J = 2\rightarrow1$ emission (Crutcher and Federman 1987) and absorption from CH and CN toward the star (Lambert et al. 1990; Crawford et al. 1994) confirmed the component structure seen by Liszt (1979).  Willson examined the correpondence between CH emission and absorption toward background stars.  Building on this correspondence, Federman and Willson (1982) used CH measurements to suggest an association between dark clouds and their envelopes of diffuse molecular gas, while Mattila (1986) showed more clearly the connection between diffuse molecular gas and dark clouds.  For the material in the direction of HD 29647, Crutcher (1985) combined radio and visible data for a comprehensive analysis of the material.  Maps of molecular emission in the vicinity of background stars were obtained to learn more about the association of diffuse and dark molecular gas.  Examples include the work of Federman and Willson (1984) on gas near the Pleiades, of Gredel et al. (1992) on gas toward HD 210121, and of Gredel et al. (1994) on translucent clouds.  The results described here are the latest effort seeking details of the correspondence revealed by the earlier studies.

Diffuse atomic and molecular gas have been traditionally studied using absorption in the ultraviolet (UV) and visible portions of the spectrum. For instance, CN absorption traces relatively dense gas (\textit{n} $\sim$ 300 cm$^{-3}$). In the portions of the cloud surrounding this gas without detectable amounts of CN absorption, we find absorption from CH, CH$^{+}\!,$ \ion{K}{1}, \ion{Ca}{1}, \ion{Ca}{2} as the density decreases. Among these species, the CN profile is usually much less complicated than profiles of other species for a given sight line because it is only probing the denser material. The column densities (\textit{N}) of CN components are correlated with column densities of corresponding CH components surrounding the denser CN regions. Welty \& Hobbs (2001) found an essentially linear relationship between \textit{N}(\ion{K}{1}) and \textit{N}(CH) by using total column densities along lines of sight. The column densities \textit{N}(\ion{Ca}{1}) and \textit{N}(\ion{K}{1}) are also well correlated (Welty et al. 2003). Figure 6 from Pan et al. (2005) shows a schematic of this idea; overlapping layered locations of tracers at visible wavelengths within a cloud. Densities (\textit{n}$_{H}$) refer to total proton density, \textit{n}(\ion{H}{1})+2\textit{n}(H$_{2}$).

\begin{itemize}
	\item \ion{Ca}{2} samples low density atomic gas (\textit{n}$_{H}$ $\leq$ 10 cm $^{-3}$)
	\item \ion{K}{1}, \ion{Ca}{1}, CH$^{+}\!,$ and CH sample somewhat denser molecular gas (\textit{n}$_{H}$ $\approx$ 100 cm $^{-3}$)
	\item CN and CO sample the densest regions of diffuse molecular clouds (\textit{n}$_{H}$  $\approx$ 300 cm $^{-3}$)
\end{itemize}

While absorption spectra are sensitive to low densities and column densities, emission spectra are sensitive to density via the collisional rate and are more difficult to detect in low density environments. Recently, it has become possible to study these environments in emission from the 158 $\mu$m fine structure line of C$^{+}$ (Langer et al. 2010).  However, because C$^{+}$ has only one fine-structure line, it cannot completely characterize the emitting gas without making assumptions about the physical conditions (Langer et al. 2014, Velusamy et al. 2010).  The current effort is meant to complement the \textit{Herschel} key program GOT C+ (Galactic  Observations of Terahertz C$^{+}$), which was designed to study the ISM by connecting the emission from the [\ion{C}{2}]  $^{2}P_{3/2}\rightarrow^{2}P_{1/2}$ fine structure line at 158 $\mu$m with emission from \ion{H}{1} at 21 cm and CO at 2.6 mm. The GOT C+ project is described in detail by Langer et al. (2010). \nocite{2010A&A...521L..18V}  The analyses of Langer et al. (2014) were based on tracers (\ion{H}{1}, $^{12}$CO, $^{13}$CO) observed at radio and millimeter wavelengths. In order to discern the types of interstellar environment producing C$^{+}$ emission, they adopted the following picture. 

 \begin{itemize}
 	\item  Diffuse Atomic Clouds (Warm Neutral Medium): Components seen in \ion{H}{1} only; the densities and column densities are too low to produce  C$^{+}$ emission above the GOT C+ sensitivity limit. (\textit{n}$_{H}$ $\leq$ 1 cm $^{-3}$)
 	\item Diffuse Atomic Clouds (Cold Neutral Medium): Components seen in \ion{H}{1},  and/or C$^{+}\!,$ with $^{12}$CO and $^{13}$CO emission well below the detection limit, with the assumption that this \ion{H}{1} is associated with the envelope around denser colder gas. Here, some of the C$^{+}$ emission comes from \ion{H}{1} and some from CO-dark H$_{2}$ gas. (1 cm $^{-3}$ $\leq$ \textit{n}$_{H}$ $\leq$ 300 cm $^{-3}$)

 	\item CO-dark H$_{2}$ Gas: Components seen in \ion{H}{1}, and/or C$^{+}\!,$ but with no $^{12}$CO and $^{13}$CO emission above the detection limit. (\textit{n}$_{H}$ $\approx$ 300 cm $^{-3}$)
 	\item Molecular Clouds: Components seen in \ion{H}{1}, and/or C$^{+}\!,$ $^{12}$CO, and $^{13}$CO emission, with the distinction that clouds with $^{13}$CO are denser. (300 cm $^{-3}$ $\leq$ \textit{n}$_{H}$ $\leq$ 1000 cm $^{-3}$)
 	
 \end{itemize}

Studies of atomic and molecular absorption and emission in the diffuse ISM have their own seemingly separate paradigms. However, measurements of atomic and molecular absorption at visible and UV wavelengths provide complementary data on some of the environments seen in the GOT C+ survey.  By associating probes seen at optical and UV wavelengths with the results from the GOT C+ survey, we can pursue an integrated study of the diffuse ISM, incorporating data from radio to UV wavelengths. We also note how these tracers are related to the CO-dark H$_{2}$ gas being probed by C$^{+}$ emission, focusing on the kinematic connections among the species. The outline of this first paper is as follows. Section 2 describes the observations. Section 3 provides the results of our survey, while Section 4 discusses connections between the various probes. The last section presents our conclusions.

\section{Observations}

\subsection{The Data}

Seventeen sight lines were chosen by examining the SIMBAD database (Wenger et al. 2000); only GOT C+ pointings accessible to the Northern Hemisphere were considered. The background targets in our survey had to be relatively bright (\textit{B}, \textit{V} $\leq$ 10) A and B stars with amounts of extinction per kpc typical for directions with diffuse molecular gas (such as that seen toward $\zeta$ Oph). The lines of sight had to be within 30 arcminutes  of the \textit{Herschel} pointings. Visual extinctions were derived by assuming $A_{V}$ = 3.1 \textit{E}(\textit{B}-\textit{V}). Distances were obtained by spectroscopic parallax as many stars did not have definitive Hipparcos or GAIA measurements and are known to approximately 20\% based on average uncertainties in magnitudes ($\sim$0.02).

The data at visible wavelengths were taken at the McDonald Observatory in December of 2012, July of 2014, and  October of 2017 with the echelle spectrograph (Tull et al. 1995). Table \ref*{obstableall} lists the stellar data and observational details for the lines of sight. References for spectral types and (B-V)$_{0}$ are provided in the footnotes of Table \ref*{obstableall}. \ion{Ca}{2} $\lambda$3933, \ion{Ca}{1} $\lambda$4226,  CH$^{+}$ $\lambda$4232, CH $\lambda$4300,  and CN $\lambda$3874 absorption features were able to all be observed with a single exposure using the 56$^{th}$ order of the echelle grating centered at 4065 \AA. This setting is denoted as \textquoteleft Blue' in Table \ref*{obstableall}. \ion{K}{1} $\lambda$7699 was observed with the  31$^{st}$ order of the echelle grating  centered on 7165 \AA. This setting is denoted as \textquoteleft Red' in Table \ref*{obstableall}.  

The data reduction was performed in IRAF using standard routines. Dark and bias frames were combined and applied to all other exposures prior to cosmic ray removal and other processing. Flat fields were normalized and divided into the exposures. A wavelength scale was established using Th-Ar lamp exposures and applied to all stellar exposures before they were corrected for their $V_{lsr}$ offset, combined, and subsequently normalized. It should be noted that \ion{K}{1} was only observed for a subset of the sight lines and has a slightly lower spectral resolution than the rest of the McDonald data, 125,000 versus 135,000. After normalization, the ISMOD routine (Sheffer, unpublished), an rms minimization spectrum synthesis FORTRAN program, was used to model the Voigt absorption profiles via spectrum synthesis and automatic rms minimization of (data minus fit) residuals.  
For each synthesized absorption feature, ISMOD fits the total column density, the number of components, relative column density fractions, radial velocity, and their Doppler widths (\textit{b}-values) for each component.   The resulting fit was compared with fits for other visible species (\ion{Ca}{2}, \ion{Ca}{1}, \ion{K}{1}, CH$^{+}\!,$ CH, CN) from the same line of sight.  If there was a disagreement in the overall component structure; all species were iteratively refit until the best overall agreement was reached.

Some species such as \ion{Ca}{2} have absorption components that extend well beyond those of other species observed at visible wavelengths. This high velocity gas seen in \ion{Ca}{2} is a well recognized phenomenon (Pan et al. 2005). This makes the determination of the extended structure more difficult. By requiring realistic \textit{b}-value ranges for each species seen in McDonald spectra (taken from the extensive survey conducted by Pan et al. 2005) and using the least number of components detected at or above three sigma, we can reduce the number of possible solutions to only secure reproducible ones. These fits were made independently of the fits to the species examined in the GOT C+ survey, and in the majority of cases, there is agreement in the structure of the components. 

Table \ref*{taboptical} gives the results of fitting the absorption from these species (ordered by Galactic coordinates given in Table \ref*{obstableall}). Although HD 35652 and HD 60146 were observed at McDonald, \textit{Herschel} measurements were not obtained, nor were data for CO or \ion{H}{1} emission. These two sight lines are not discussed further. Here we focus on velocities of components; the column densities are needed for the analyses in the subsequent paper.

The GOT C+ project is described in detail by Langer et al. (2010) and Langer et al. (2014). \nocite{2010A&A...521L..18V} GOT C+ sources are labeled as GXXX.X+YY which are the latitude and longitude rounded to one decimal, the actual coordinates are given in Table \ref*{Herschellos}. The details of the data reduction are described in Pineda et al. (2013). The GOT C+ data sets are available as a \textit{Herschel} User Provided Data Product\footnote[1]{http://www.cosmos.esa.int/web/herschel/user-provided-data-products}. The GOT C+ data were fit independently of the results from the McDonald data. GOT C+ data were fitted with 1-3 Gaussians on top of a linear continuum, in IDL, using the MPFIT package from Markwardt et al. (2009). The initial fits for the Gaussians were determined by eye.

The [C II] $^{2}$P$_{3/2}\rightarrow^{2}$P$_{1/2}$ observations from the HIFI (Pilbratt et al. 2010; de Graauw et al. 2010) instrument on-board the \textit{Herschel} space observatory have an angular resolution of 12\arcsec. The observations of the J = 0 $\rightarrow$ 1 transitions of $^{12}$CO, $^{13}$CO, and C$^{18}$O from the ATNF Mopra Telescope have an angular resolution of 33\arcsec.  The \ion{H}{1} 21 cm observations from the VLA Galactic Plane Survey (VGPS; Stil et al. 2006) have an angular resolution of 1\arcmin. The species observed in emission C$^{+}\!,$ CO, and \ion{H}{1} have a velocity resolution of 0.8 km s$^{-1}$ while C$^{18}$O has a velocity resolution of 1.6 km s$^{-1}\!.$ The species observed in absorption (\ion{Ca}{2}, \ion{Ca}{1}, \ion{K}{1}, CH$^{+}\!,$ CH, CN) have a velocity resolution of 2.2 km s$^{-1}\!.$ Each detection of a component is required to be at least at the 3 sigma level. The uncertainty in column density for each component is between 1\% and 30\% of the column density of that component. Only the smallest components have uncertainties nearing the three sigma detection limit; typical total uncertainties are closer to 10\%.  The uncertainty in velocity is $\sim$0.3 km s$^{-1}$ for strong lines and $\sim$0.6 km s$^{-1}$ for weak lines, while the typical uncertainty in \textit{b}-value is also $\sim$0.3 km s$^{-1}\!$.

Each sight line has a stacked plot showing the spectra of the species that were observed in absorption toward the star and in emission for the closest GOT C+ pointing.  Species not shown were not observed for that pointing, with the exception of C$^{18}$O which is not shown due to the low correspondence with absorption components. The $V_{lsr}$ of components detected at a three sigma level are indicated for each species as a red tick above the spectra. Our results of the absorption and emission are combined into Tables \ref*{outertab} and \ref*{innertab}. These two tables show the $V_{lsr}$ for components with emission features and the $V_{lsr}$ of any corresponding absorption features for the inner and outer Galaxy, respectively. These components, seen in emission and absorption, are discussed in Section 3. However, there are often components seen in absorption that do not correspond with any of the components seen in emission, or vice versa. 

The following points were considered when seeking matches to component velocities.  First, diffuse atomic and molecular gas, with kinetic temperatures of 50 to 80 K, have thermal widths of about 1 km s$^{-1}$.  The extensive survey conducted by Pan et al. (2005) revealed typical \textit{b}-values of 1 to 2 km s$^{-2}$ for absorption lines, depending on species. Since the species in our survey from the McDonald observations have atomic masses greater than 20 amu, thermal broadening makes a negligible contribution to the b-value, which mainly arises from turbulent motion after removing the instrumental width of 2.2 km s$^{-1}$.  Second, the light path for the hollow cathode lamp used in wavelength calibration differs slightly from that of the stellar radiation.  This difference causes 0.5 to 1.0 km s$^{-1}$ offsets from an absolute velocity scale.  Third, most of the background stars in our survey have distances within 500 pc of the Sun, with only one significantly beyond 1000 pc.  On Galactic scales, these stars are considered within the solar neighborhood.  The majority of the gas along the lines of sight have velocities in the Local Standard of Rest less than 10 km s$^{-1}$.  Dynamical phenomena like stellar winds and expanding supernova remnants create components with larger radial velocities, some of which are in excess of 100 km s$^{-1}$.  An example involves directions that probe the supernova remnant IC 443 (Welsh and Sallman 2003; Hischauer et al. 2009).  This leads to complications in associating components seen in absorption and emission in the inner Galaxy where large velocities may arise from dynamical phenomena, especially in atomic gas, or Galactic rotation. In light of these considerations, correspondences were noted when velocities associated with line centers agreed within the full width at half maximum of the lines.  Weighted averages based on column density were performed for some species when comparisons were made with the broadest lines (usually H~{\small I}).

\subsection{Emission versus Absorption Considerations}

When comparing observations in absorption and in emission, there will frequently be components that appear shifted, blended, or even missing entirely. Over half (56\%) of the emission components are not associated with an absorption component, and over half (53\%) of the absorption components are not associated with an emission component. The majority of the unassociated components are only seen in species associated with low density gas. Over half (52\%) of the unassociated emission components are only seen in \ion{H}{1} and 80\% of the unassociated absorption components are only seen in \ion{Ca}{2}.

There are several sources of these inconsistencies. One is the fact that absorption lines only sample gas along the line of sight between the background source and the observer, while the emission lines sample all the gas. This is the reason components observed in emission at high $V_{lsr}$ are sometimes absent in absorption.  The $V_{lsr}$ structure for two nearby absorption sight lines can exhibit similar differences when the background sources are at different distances.

Secondly, at low densities almost all the gas seen in absorption is in the lowest states and the maximum column density is probed while emission lines typically come from weakly excited states in diffuse gas and are sensitive to density. This difference results in the absorption measurements being more sensitive to detecting the gas than the emission lines, but is also restricted to lines of sight with background sources while emission lines can be mapped. This is one of the sources for additional components seen in absorption that are not seen in emission. Over half (80\%) of the unassociated absorption components are only seen in \ion{Ca}{2}, which indicates low density gas.

Another source of inconsistency is from the blending of lines. Sometimes the relatively broad emission lines result in a component that corresponds to a blend of two absorption components with slightly different velocities. Whenever two components had to be combined to form a corresponding component, it is indicated by the use of the superscript \textquoteleft a' on the $V_{lsr}$ in Tables \ref*{outertab} and \ref*{innertab}, where an averaged velocity (weighted by column density) is quoted for the McDonald data.

There are occasional additional components seen in emission that are not seen as absorption components and fall within the overlapping $V_{lsr}$ ranges. The most likely explanation is from variations in the small scale structure between the different locations of the emission and absorption measurements. In general, whenever an observed species is associated with one or more of the unobserved species, those unobserved species are likely present at varying levels between the emission and absorption pointings. However, at the location of the measurements, the levels are too low to be confidently measured. Table \ref*{G010.4+0.0table} shows a summarized component-by-component interpretation for a pointing with two nearby sight lines. The majority of components for the two sight lines are the same, but there are also clear differences. The component with a $V_{lsr}$ of 7 km s$^{-1}$ is seen in emission toward the GOT C+ pointing and in absorption toward HD 165783 but not toward the other nearby slight line, HD 165918.  This small scale structure can also cause components to be slightly shifted in velocity.

\section{Results}

To illustrate the general results, a component-by-component analysis is shown for G014.8-1.0 (see Table \ref*{G014.8table}). The remainder of the sight lines are discussed in terms of connections to nearby dark clouds and notable deviations from the general results. Upper limits for the linear separation are calculated using the derived distance to the target star and the observed angular separation; if two targets have the same angular separation, the distance to the nearest target is used in the calculations.

\subsection{A Component by Component Discussion \\ for G014.8-1.0}
HD 168607 is at a distance of 1100 pc and is 12 arcmin away ($\leq$ 3.8 pc at 1100 pc) from the pointing G014.8-1.0. Almost all the components seen in \ion{H}{1} are also seen in \ion{Ca}{2}, indicating they are both probes of diffuse regions of the individual clouds. However, there are many components seen in \ion{Ca}{2} that are not seen in emission (at $V_{lsr}$ of -45.5, -39.7, -33.6, -28.4, -18.2, -13.4, -5.3, -0.8,  3.4, 9.9, 59.2, 65.6 km s$^{-1}$). This difference is due to the increased sensitivity of absorption measurements relative to emission measurements. There are also some \ion{H}{1} features that do not have corresponding absorption features; these features have large $V_{lsr}$ (70.0 and 112.6 km s$^{-1}$) and are most likely probing the material behind HD 168607. 

There are three components that are seen in \ion{Ca}{2}, CH$^{+}\!,$ and CH (near $V_{lsr}$ of 6.7, 15.0, 23.3 km s$^{-1}$); these components only correspond with emission features seen in \ion{H}{1}. The presence of CH and CH$^{+}$ indicates that these components are probing denser regions than if only \ion{Ca}{2} or \ion{H}{1} were observed. We would expect these types of regions, with CH or CH$^{+}$ detections, to have some amount of CN or CO. The CH$^{+}$ absorption indicates the presence of relatively low density gas (Pan et al. 2005); otherwise, reactions with H$_{2}$ would destroy CH$^{+}\!.$ There are clear relationships involving CO, CH$^{+}\!$, and CH with with molecular hydrogen for \textit{N}(H$_{2}$)$\geq$10$^{19}$ cm$^{-2}$, as documented in Sheffer et al. (2008). These H$_{2}$ column densities correspond to molecular hydrogen fractions of 5\% to 10\%. This is the minimum molecular hydrogen fraction we use to define gas as molecular. In low density gas, the CH$^{+}$ + O $\rightarrow$ CO$^{+}$ + H reaction leads to CO production. The CH and CO are also coupled through the neutral-neutral reaction route CH + O $\rightarrow$ CO + H. However, the abundances of these species are not high enough for a reliable detection in CO emission.  Therefore, these components are most likely probing CO-dark H$_{2}$ regions, where CO is present but at a column density too low to be confidently measured in emission.    

There is one component with detectable CN, CH, and \ion{Ca}{2} near a $V_{lsr}$ of 18.7 km s$^{-1}$. This component corresponds to emission in C$^{+}\!,$ $^{12}$CO, $^{13}$CO, and C$^{18}$O. The presence of $^{12}$CO, $^{13}$CO, and C$^{18}$O all indicate that this component is probing much denser gas. Here, the associated data from absorption are revealing the diffuse outer regions of a molecular cloud.

There are several components seen in CO emission that are not seen as molecular absorption components (at $V_{lsr}$ of -26.1, 30.0, 39.1, 46.3, 55.9, 75.6 km s$^{-1}$), despite the majority being associated with components in \ion{Ca}{2}. The presence of CO emission indicates that the density in the area being probed should be high enough for detectable amounts of other species, such as CH or CH$^{+}\!.$ There will be some differences that arise from variations in the small scale structure between the different locations of the emission and absorption measurements, in this case 12 arcmin. Thus, the presence of only \ion{Ca}{2} indicates that the absorption measurements are likely probing the lower density atomic region that surrounds the molecular gas associated with the CO emission. 
Table \ref*{G014.8table} provides a summarized component-by-component interpretation for this sight line.  See Figure 3 for the stacked spectra.

\subsection{Summaries for Other Pointings}

Tables \ref*{outertab} and \ref*{innertab} show the $V_{lsr}$ structure of components with emission features and the $V_{lsr}$ of any corresponding absorption features for the inner and outer Galaxy. However, there are often components seen in absorption that do not correspond with any of the components seen in emission. These components are discussed here.

\bigskip
\noindent\textbf{G010.4+0.0}: 
HD 165783 is 4 arcmin away and HD 165918 is 6 arcmin away from G010.4+0.0. Assuming a distance of 500 pc, the separations between directions related to the absorption and emission measurements are $\leq$ 0.6 pc and $\leq$ 0.9 pc, respectively. 

 There are five components seen in emission and toward both HD 165783 and HD 165918 (near $V_{lsr}$ of -21.2, -10.3, 0.0, 4.2, 10.7 km s$^{-1}$). Three of these are only associated with \ion{H}{1} emission and are probing the lower density regions (near $V_{lsr}$ of -21.2, 0.0, 10.7 km s$^{-1}$). One component, seen in \ion{Ca}{2} at 4.2 km s$^{-1}\!,$ is also associated with CH/CH$^{+}$ absorption and C$^{+}$ emission. The presence of CH and CH$^{+}$ without CO indicates this is likely probing diffuse molecular gas and CO-dark  H$_{2}$ gas.

There are six components from species seen in absorption without corresponding emission features. These additional components are seen toward HD 165783 near $V_{lsr}$ (km s$^{-1}$) of -15.9 (\ion{Ca}{2} and CH$^{+}$), -6.4 (\ion{Ca}{2}), -3.6 (\ion{Ca}{2} and CH$^{+}$), and 13.5, 16.6, 39.0 (\ion{Ca}{2}). Only three of these additional components are seen toward HD 165918 near $V_{lsr}$ (km s$^{-1}$) of -16.0, -3.5, 16.0 (\ion{Ca}{2}). These differences likely stem from small scale variations in structure and weak \ion{Ca}{2} components. There are several nearby dark clouds including ZHS2005 C3C, C3B, C3A (Zhang et al. 2005) but all measured $V_{lsr}$ are beyond the observed range spanned by the sight lines toward HD 165783 or HD 165918.	The $V_{lsr}$ given in the literature (Zhang et al. 2005) for C$^{18}$O at 66 km s$^{-1}$ agrees with our measurements of a component seen in C$^{18}$O, C$^{+}\!,$ and \ion{H}{1} at 65 km s$^{-1}\!.$  Table \ref*{G010.4+0.0table} shows the summarized component-by-component interpretation for this GOT C+ pointing for both of its two nearby stellar sight lines. See Figures 1 and 2 for the stacked spectra. The structure of each individual component can be seen in Figure 1 as dotted lines with the center of each component indicated by a red tick mark. Other stacked spectra indicate the components with only red the tick marks as in Figure 2.

\bigskip
\noindent\textbf{G015.7+1.0}: HD 167498 is 9 arcmin away and HD 167812 is 15 arcmin away. Assuming a distance of 250 pc, the separations between the directions observed in absorption and emission are $\leq$ 0.7 pc and $\leq$ 1.1 pc, respectively.  There is one component seen in emission from G015.7+1.0 with \ion{H}{1} and toward both HD 167498 and HD 167812 in \ion{Ca}{2} and CH (near a $V_{lsr}$ of $\sim$4.1 km s$^{-1}$). This component is likely probing CO-dark H$_{2}$  gas. There are two weak \ion{Ca}{2} components toward HD 167498 that have no emission features associated with them (near $V_{lsr}$ of -14.7 and -10.8 km s$^{-1}$). There is also one component toward HD 167498 and HD 167812 observed in absorption from CH, CH$^{+}\!,$ and \ion{Ca}{2} that has no emission feature (near a $V_{lsr}$ of 1.0 km s$^{-1}$). This component is likely probing CO-dark H$_{2}$ gas. 
See Figures 4 and 5 for the stacked spectra.

\bigskip
\noindent\textbf{G020.0+0.0}: 
HD 169754 is 14 arcmin away from G020+0.0. Assuming a distance of 2000 pc, the directions are $\leq$ 8.1 pc apart.  Almost all the \ion{H}{1} emission components are also seen in \ion{Ca}{2}. 
 There are two components from species detected via absorption toward HD 169754 that are unassociated with components seen in emission near $V_{lsr}$ (km s$^{-1}$) of -7.3 (\ion{Ca}{2}) and 12.2 (\ion{Ca}{2} and CH$^{+}$). There are several nearby dark clouds including L413 (Lynds 1962) but only one with additional velocity information. BGPS G020.050+0.195 has a velocity component around 18 km s$^{-1}$ seen in HCO$^{+}\!,$ N$_{2}$H$^{+}\!,$ and NH$_{3}$ (Schlingman et al. 2011) that agrees with a component seen in \ion{Ca}{2}, \ion{H}{1} and CH$^{+}$ near 20 km s$^{-1}\!,$ and is likely associated with the outer envelope of the molecular cloud. 
See Figure 6 for the stacked spectra.

\bigskip	
\noindent\textbf{G032.6+0.0}: 
HD 174509 is 13 arcmin away from G032.6+0.0 and assuming a distance of 500 pc, the directions probe material $\leq$ 1.9 pc apart. The component with \ion{H}{1} emission at -0.5 km s$^{-1}$ is associated with \ion{Ca}{2}, \ion{Ca}{1}, \ion{K}{1}, \ion{CH$^{+}$}{0} and \ion{CH}{0} absorption and is likely probing diffuse molecular gas.  The component with \ion{H}{1} and \ion{C$^{+}$}{0} emission near 14.4 km s$^{-1}$ is associated with \ion{Ca}{2}, \ion{Ca}{1} indicating diffuse atomic gas.  The other three absorption components all have detected \ion{Ca}{2}, \ion{Ca}{1}, \ion{K}{1},  \ion{CH}{0} and \ion{CN}{0} and are associated with CO emission near $V_{lsr}$ of 5.8, 9.7, and 12.1 km s$^{-1}$ and are likely probing the cloud envelope. 

There are several dark clouds nearby with measurements in NH$_{3}$: BGPS G032.524-00.131 (79.3 km s$^{-1}$), BGPS G032.527-00.123 (79.4 km s$^{-1}$), BGPS G032.411-00.0099 (80.2 km s$^{-1}$), BGPS G032.436-00.17 (77.4 km s$^{-1}$) (Dunham et al. 2011); however all the corresponding $V_{lsr}$ measurements lie beyond the range measured for species in absorption. There is agreement with a component seen in emission in $^{12}$CO, $^{13}$CO, C$^{18}$O, C$^{+}\!,$ and \ion{H}{1} near 75 km s$^{-1}\!,$ indicating this molecular cloud lies behind HD 174509. See Figure 7 for the stacked spectra.

\bigskip
\noindent\textbf{G091.7+1.0}: 
BD+49$^{\circ}$ 3482 is 9 arcmin away and BD+49$^{\circ}$ 3484 is 4 arcmin away from G091.7+1.0. For a distance of 250 pc, the directions are $\leq$ 0.7 pc and $\leq$ 0.3 pc apart, respectively. It should be noted that the pointing G091.7+1.0 did not include CO measurements. There is one component seen in \ion{H}{1} emission and in \ion{Ca}{2} and  CH$^{+}$ for both BD+49$^{\circ}$ 3482 and BD+49$^{\circ}$ 3484 near $V_{lsr}$ of 3.5 km s$^{-1}\!.$ There is also a component near 1.6 km s$^{-1}$ seen toward BD+49$^{\circ}$ 3482 that is associated with K I and CH$^{+}$ absorption and \ion{H}{1} emission. 

The weak \ion{Ca}{2} component from BD+49$^{\circ}$ 3482 (at $V_{lsr}$ of -4.4 km s$^{-1}$) has no emission features associated with it and three \ion{Ca}{2} components toward BD+49$^{\circ}$ 3484 also have no associated emission features (at $V_{lsr}$ of 8.5, 10.3 and 13.0 km s$^{-1}$). Two of the components from BD+49$^{\circ}$ 3484 also have a CH component (near 8.5 and 10.3 km s$^{-1}$). One of these components also has CN absorption (near 10.3 km s$^{-1}$). 
There are two dust clouds along adjacent directions, Dobashi 2934 (Dobashi 2011) and TGU H541 P30 (Dobashi et al. 2005).  There is also a molecular cloud, [DBY 94] 091.7+00.9 observed in $^{13}$CO with $V_{lsr}$ of 10.9 km s$^{-1}$ (Dobashi et al. 1994).  This agrees nicely with the absorption seen in  CN (10.2 km s$^{-1}$),  CH (10.6 km s$^{-1}$), and \ion{Ca}{2} (10.3 km s$^{-1}$) and C$^{+}$ emission, indicating the presence of a diffuse molecular envelope.  The nearby sight line (BD+49$^{\circ}$ 3482), although farther from the \textit{Herschel} pointing, mainly shows the lower velocity material.
See Figures 8 and 9 for the stacked spectra.	

\bigskip
\noindent\textbf{G109.8+0.0}: 
HD 240179 and HD 240183 are both 7 arcmin away from G109.8+0.0 and have a separation of $\leq$ 1.4 pc, assuming a distance of 700 pc.  It should be noted that the pointing G109.8+0.0 did not include CO measurements. There are two components seen in \ion{H}{1} emission and  \ion{Ca}{2} absorption toward both HD 240179 and HD 240183 (near $V_{lsr}$ of -10.3 and 0.2 km s$^{-1}$). Spectra of HD 240179 also reveal this component in CH$^{+}\!,$ as well as an additional component near 4.0 km s$^{-1}$ that is not seen toward HD 240183. This component near 4.0 km s$^{-1}$ is detected in \ion{H}{1} emission and \ion{Ca}{2} and K I absorption and is likely sampling a diffuse molecular cloud edge. 

There are several components seen in absorption that have no corresponding emission features (near $V_{lsr}$ of -7.0, -3.5, -1.0, 7.6, 11.9 and 17.5 km s$^{-1}$), several of which are only weak \ion{Ca}{2} components (near $V_{lsr}$ of 7.6, 11.9, and 17.5 km s$^{-1}$). However, there are three components with \ion{Ca}{2}, CH, CH$^{+}\!,$ and K I with no corresponding emission features (near $V_{lsr}$ of -7.0, -3.5, -1.0 km s$^{-1}$) and might point to the presence of CO-dark H$_{2}$ gas, but without CO emission data for this pointing it is difficult to confirm.
There is a dense core nearby, IRAS 23033+5951, which has been studied extensively.  NH$_{3}$ observations by Wouterloot et al. (1988) show emission at $V_{lsr}$ of -52.8 km s$^{-1}\!,$ while $^{12}$CO and $^{13}$CO measurements by Wouterloot et al. (1989) indicate respective $V_{lsr}$ of -51.8 and -53.1 km s$^{-1}\!.$  This component is clearly associated with the emission seen in \ion{H}{1} and C$^{+}\!.$  
See Figures 10 and 11 for the stacked spectra.

\bigskip
\noindent\textbf{G207.2-1.0}: 
HD 47073 and HD 260737 are both 7 arcmin away from G207.2-1.0, and assuming a distance of 200 pc, the directions are both separated by $\leq$ 0.4 pc. There is one component seen in \ion{H}{1} emission and in \ion{Ca}{2} absorption for both HD 47073 and HD 260737 near a $V_{lsr}$ of 3.8 km s$^{-1}$ and is likely only probing diffuse atomic gas. There is one additional component toward HD 260737 that is seen in both emission (H I and $^{12}$CO) and absorption (\ion{Ca}{2} and K I) near a $V_{lsr}$ of 9.4 km s$^{-1}$ and is likely probing the envelope of a molecular cloud. Four additional weak \ion{Ca}{2} components toward HD 260737 are not seen in emission (-15.6, -8.6, -2.9, 1.8, and 5.3 km s$^{-1}$). 
See Figures 12 and 13 for the stacked spectra.

\bigskip
\noindent\textbf{G225.3+0.0}: 
HD 55469 is 26 arcmin away and HD 55981 is 8 arcmin away from G225.3+0.0. Assuming a distance of 400 pc, the separations are $\leq$ 3.0 pc and $\leq$ 0.9 pc, respectively. There is no \ion{H}{1} data for this pointing. Emission components are detected toward G225.3+0.0 near $V_{lsr}$ of 17.8 (C$^{+}$ and $^{12}$CO) and 51.9 (C$^{+}$) km s$^{-1}\!.$ There are absorption components detected toward HD 55981 near $V_{lsr}$ of -7.3, -2.3, 2.8, 5.7, 9.2 (\ion{Ca}{2}), 14.8 (\ion{Ca}{2}, \ion{K}{1}, and CH$^{+}$), and 20.9 km s$^{-1}$ (\ion{Ca}{2}). There are also absorption components detected toward HD 55469 near $V_{lsr}$ of 2.0 (\ion{Ca}{2} and \ion{K}{1}) and 3.8, 10.3 km s$^{-1}$ (\ion{Ca}{2}). While there are components seen in emission and in absorption, no clear correspondence is found between the emission and absorption components. The closest correspondence is seen for the absorption from CH$^{+}\!,$ \ion{K}{1} and \ion{Ca}{2} around a $V_{lsr}$ of 14.8 km s$^{-1}$ and the emission from C$^{+}$ and $^{12}$CO around a $V_{lsr}$ of 17.5 km s$^{-1}\!.$ Kim et al. (2004) studied L1658 in $^{13}$CO at coordinates G224.3-1.1 (at a $V_{lsr}$ of 15.5 km s$^{-1}$) and G226.1-0.04 (at a $V_{lsr}$ of 16.2 km s$^{-1}$). These velocities both fall between those of the emission at 17.5 km s$^{-1}$ and the absorption at 14.8 km s$^{-1}\!,$ suggesting that the components are related. See Figures 14 and 15 for the stacked spectra.

\section{Discussion} 
Section 3.1 gave a component-by-component analysis for a single sight line and Tables \ref*{G010.4+0.0table} and \ref*{G014.8table}  show component-by-component interpretations for single pointings. Section 3.2 provided similar results in summary. When the results of all the pointings are combined, trends start to emerge. Table \ref*{Sumres7} shows the tabulated results for all observations and we interpret the combined results here.

\subsection{\ion{H}{1} and  \ion{Ca}{2} Detections}

The majority of components seen in \ion{H}{1} are also seen in \ion{Ca}{2}, which are both probes of the more diffuse regions of individual clouds. 
Components only seen in \ion{Ca}{2} and \ion{H}{1} gas are likely probing diffuse atomic gas, where densities are too low to excite [C II] $^{2}$P$_{3/2}\rightarrow^{2}$P$_{1/2}$ efficiently or react rapidly enough to produce CO. This accounts for approximately 1/4 of the associated components.
Components seen in \ion{Ca}{2} and \ion{H}{1} gas that are also associated with any of the additional species considered here typically probe denser gas than the diffuse atomic regime.

\subsection{C$^{+}\!,$ \ion{Ca}{1}, \ion{K}{1}, CH$^{+}\!,$ CH, and CN Detections }

The majority of components seen in 
C$^{+}\!,$ \ion{Ca}{1}, \ion{K}{1}, CH$^{+}\!,$ CH, or CN also tend to be associated with probes of the very diffuse gas (\ion{H}{1} and \ion{Ca}{2}). The presence of C$^{+}\!,$ \ion{Ca}{1}, \ion{K}{1}, CH$^{+}\!,$ or CH indicates that the component is probing denser material than if only \ion{Ca}{2} or \ion{H}{1} were observed (Pan et al. 2005). The additional presence of  CN indicates that the component is probing even denser material than if only C$^{+}\!,$ \ion{Ca}{1}, \ion{K}{1}, CH$^{+}\!,$ or CH were observed.

The species that probe very diffuse gas in emission seem to be associated with species seen in absorption that are seen in diffuse molecular clouds. The presence of C$^{+}\!,$ \ion{Ca}{1}, \ion{K}{1}, CH$^{+}\!,$ or CH all suggest that there should be some CO being produced, as discussed in Section 3.1, and the presence of CN suggests even higher levels of CO should be present. Thus if there is no detectable amounts of CO at the same $V_{lsr}$, it does not mean CO is not present, but that the conditions and/or density are not suitable for detecting the emission. Therefore, these components with C$^{+}\!,$ \ion{Ca}{1}, \ion{K}{1}, CH$^{+}\!,$ or CH and no detectable CO are most likely probing regions of CO-dark H$_{2}$ gas and components associated with CN and no detectable CO are probably probing the slightly denser inner regions of CO-dark H$_{2}$ gas.    

These regions of CO-dark H$_{2}$ gas can be further probed by looking at the pure rotational emission lines of H$_{2}$. These lines have been detected by ISO and then by \textit{Spitzer} several times, e.g. the Taurus molecular cloud (Goldsmith et al. 2010) and in translucent clouds (Ingalls et al. 2011). JWST would likely build on these results using the rotational emission lines of H$_{2}$ to probe CO-dark H$_{2}$ gas at the edges of molecular clouds.    

\subsection{CO Detections}

Detectable amounts of CO emission often correspond to components seen in C$^{+}\!,$ \ion{K}{1}, CH$^{+}\!,$ CH, or CN. This accounts for almost half of the associated CO components. The other half of the associated CO emission components are only associated with \ion{Ca}{2} components. About 70\% of the CO emission components also 
have $^{13}$CO or C$^{18}$O emission. The additional presence of $^{13}$CO or C$^{18}$O indicates a component that is associated with much denser gas, specifically the inner portion of the associated molecular cloud. Detectable amounts of C$^{18}$O are only present in the most dense gas regions considered here.

\section{Conclusions}

The majority of absorption components are only associated with gas seen in \ion{H}{1} emission. About half of these components are representative of diffuse atomic gas, probed by \ion{Ca}{2}, \ion{Ca}{1}, and \ion{K}{1}. The amount of C$^{+}$ in atomic gas from the Warm Neutral Medium is usually too low to produce C$^{+}$ emission above the GOT C+ sensitivity limit. However, in the atomic and molecular gas of the Cold Neutral Medium, C$^{+}$ begins to increase above the GOT C+ sensitivity limit. 

More importantly, the remaining half of the absorption components associated with \ion{H}{1} only and/or C$^{+}$ gas come from molecules found in diffuse molecular gas (e.g., Pan et al. 2005). In diffuse gas, reactions between C$^{+}$ and H$_{2}$ lead to CH$^{+}$ and CH, and further reactions lead to CN (e.g., Federman et al. 1994) and CO (e.g., Sheffer et al. 2008), despite the amount of emission being below the sensitivity limit. Therefore, components with C$^{+}\!,$ Ca I, \ion{K}{1}, CH$^{+}\!,$ or CH and no CO emission are likely associated with CO-dark H$_{2}$ gas.

There were very few detections of CN absorption in this sample, but when CN is seen, it is usually connected with gas containing CO emission. This further supports the combined picture summarized below. Furthermore, whenever these absorption components are associated with CO gas seen in emission, they are most likely probing the molecular cloud envelope. In the densest regions probed here, we also start to detect $^{13}$CO and C$^{18}$O emission. In summary, we can say

\begin{itemize}

\item Diffuse Atomic Clouds have components seen in: \ion{Ca}{2}, \ion{Ca}{1}, \ion{H}{1}, and sometimes C$^{+}\!;$ (\textit{n}$_{H}$ $<$ 100 cm $^{-3}$)

\item Diffuse Molecular Clouds (CO-dark H$_{2}$ gas) have components seen in: C$^{+}\!,$ \ion{K}{1}, CH$^{+}\!,$ CH, or CN but without $^{12}$CO or $^{13}$CO emission; (100 cm $^{-3}$ $\leq$ \textit{n}$_{H}$ $\leq$ 300 cm $^{-3}$)

\item Dense Molecular Cloud Envelopes have components seen in: $^{12}$CO and $^{13}$CO emission (with the distinction that clouds with $^{13}$CO are denser) and any of the additional species seen in emission or absorption. (300 cm $^{-3}$ $\leq$ \textit{n}$_{H (envelope)}$ $\leq$ 1000 cm $^{-3}$)

\end{itemize}

This combined picture of diffuse atomic and molecular gas will be developed further in a subsequent paper, where analyses of CN chemistry as well as C$^{0}\!,$ CO, and CN excitation yield estimates for density. The data for \ion{C}{1} and CO come from recently acquired \textit{Hubble Space Telescope} measurements.

\acknowledgments
We are grateful to Yaron Sheffer for allowing us the use of his program ISMOD. This research made use of the SIMBAD database operated at CDS, France. Some of this work was preformed at the Jet Propulsion Laboratory, California Institute of Technology, under contract with NASA. U.S. Government sponsorship acknowledged.

\software{IRAF, ISMOD, MPFIT (Markwardt et al. (2009)}


\section{Tables and Figures}
\clearpage									

\begin{deluxetable}{lccccccccc}[h]
	
	\tablecaption{Stellar Data \label{obstableall}}
	\tablehead{\colhead{Star}  & \colhead{\textit{l} }  & \colhead{\textit{b}}  & \colhead{\textit{B}}  & \colhead{ \textit{V} } & \colhead{\textit{E}(\textit{B}-\textit{V})}  &\colhead{Distance}  &\colhead{Spectral}  & \colhead{Exp time}& \colhead{Exp time}\\
		& \colhead{}&\colhead{} & \colhead{(Mag)}& \colhead{(Mag)} & \colhead{(Mag)}& \colhead{(pc)} &\colhead{Type} & \colhead{Blue (sec)}& \colhead{Red (sec)} }
	
	\startdata
	\\
\multicolumn{10}{c}{\underline{July 2014 Observations}} \\																			
\\																			
HD 165783	&	10.45	&	0.06	&	8.60\tablenotemark{a}	&	8.33\tablenotemark{a}	&	0.46	&	500	&	B3/5Ib\tablenotemark{b}         	&	5400	&	$\ldots$	\\
HD 165918	&	10.51	&	-0.07	&	8.22\tablenotemark{c}	&	8.11\tablenotemark{c}	&	0.26	&	900	&	B5IV/V\tablenotemark{b}         	&	4800	&	$\ldots$	\\
HD 168607	&	14.97	&	-0.94	&	9.82\tablenotemark{d}	&	8.28\tablenotemark{d}	&	1.37	&	1100	&	B9Iaep\tablenotemark{e}         	&	7200	&	$\ldots$	\\
HD 167498	&	15.65	&	0.85	&	8.07\tablenotemark{c}	&	7.83\tablenotemark{c}	&	0.19	&	400	&	A2IV\tablenotemark{b}           	&	3600	&	$\ldots$	\\
HD 167812	&	15.88	&	0.59	&	8.19\tablenotemark{c}	&	7.97\tablenotemark{c}	&	0.29	&	250	&	B9.5V\tablenotemark{b}          	&	3600	&	$\ldots$	\\
\enddata
	\tablenotetext{a}{Jaschek and Egret 1982}
	\tablenotetext{b}{Houk and Smith-Moore 1988}
	\tablenotetext{c}{Hog et al. 2000}
	\tablenotetext{d}{Ducati 2002}
	\tablenotetext{e}{Walborn and Fitzpatrick 2000}
	\tablenotetext{f}{Morgan et al. 1955}
	\tablenotetext{g}{Cannon and Pickering 1993}
	\tablenotetext{h}{Houk and Swift 1999}
	\tablenotetext{i}{ESA 1997}	\tablenotetext{j}{Kislyuk et al. 2000}
	\tablenotetext{k}{Hohle et al. 2010}
	\tablenotetext{l}{Guetter 1968}
	\tablenotetext{m}{ Claria 1974}
\end{deluxetable}
\clearpage
\newpage

\begin{longrotatetable}
	\begin{deluxetable}{cccccccccccccccccccccccc}
		\tabletypesize{\tiny}
		\setlength{\tabcolsep}{3pt}
		\tablecaption{Observations from McDonald Observatory \label{taboptical}}		
		\tablehead{	
			\colhead{ Star }                             &\colhead{ } & \colhead{Ca II} & 
			& \colhead{ } &\colhead{ } & \colhead{K I} & \colhead{ }
			& \colhead{ } &\colhead{ } & \colhead{Ca I} & \colhead{ }
			& \colhead{ } &\colhead{ } &\colhead{CH$^{+}$} &\colhead{ }
			& \colhead{ } &\colhead{ } &  \colhead{CH} &\colhead{ }
			& \colhead{ } &\colhead{ } & \colhead{CN}
			& \colhead{ } 
			\\
			\cline{2-4}  \cline{6-8}  \cline{10-12} \cline{14-16} \cline{18-20} \cline{22-24}
			\colhead{ }                             &\colhead{ $V_{lsr  }$ } & \colhead{\textit{N}/10$^{11}$} & \colhead{\textit{b} }
			&\colhead{ }&\colhead{ $V_{lsr}$} &  \colhead{\textit{N}/10$^{11}$} & \colhead{\textit{b} }
			&\colhead{ }&\colhead{ $V_{lsr  } $} & \colhead{\textit{N}/10$^{11}$} & \colhead{\textit{b} } &\colhead{  }                            &\colhead{ $V_{lsr }$} & \colhead{\textit{N}/10$^{11}$} & \colhead{  \textit{b} }
			&\colhead{ }&\colhead{ $V_{lsr }$ } & \colhead{\textit{N}/10$^{11}$} & \colhead{\textit{b} }
			&\colhead{ }&\colhead{ $V_{lsr }$ } & \colhead{\textit{N}\tablenotemark{a}/10$^{11}$} & \colhead{\textit{b} }
			\\ 
			\colhead{ } &\colhead{(km s$^{-1}$) } & \colhead{(cm$^{-2}$)} & \colhead{(km s$^{-1}$)} &\colhead{ }&\colhead{(km s$^{-1}$)} & \colhead{(cm$^{-2}$)} & \colhead{(km s$^{-1}$)} &\colhead{ }&\colhead{(km s$^{-1}$)} & \colhead{(cm$^{-2}$)} & \colhead{ (km s$^{-1}$)} &\colhead{ }&\colhead{(km s$^{-1}$)} & \colhead{(cm$^{-2}$)} & \colhead{(km s$^{-1}$)}  &\colhead{ }&\colhead{(km s$^{-1}$)} & \colhead{(cm$^{-2}$)} & \colhead{(km s$^{-1}$)}	&\colhead{ }&\colhead{(km s$^{-1}$)} & \colhead{(cm$^{-2}$)} & \colhead{(km s$^{-1}$)} 
		}
		\startdata
		HD 165783	&	-21.5	&	0.6	&	2.5	&	&	$ \ldots$	&	$ \ldots$	&	$ \ldots$ 	&	&	$ \ldots$	&	$ \ldots$	&	$ \ldots$	&	&	$ \ldots$	&	$ \ldots$	&	$ \ldots$	&	&	$ \ldots$	&	$ \ldots$	&	$ \ldots$	&	&	$ \ldots$	&	$ \ldots$	&	$ \ldots$	\\
		&	-15.9	&	2.4	&	2.0	&	&	$ \ldots$	&	$ \ldots$	&	$ \ldots$ 	&	&	$ \ldots$	&	$ \ldots$	&	$ \ldots$	&	&	-15.3	&	58.0	&	2.4	&	&	$ \ldots$	&	$ \ldots$	&	$ \ldots$	&	&	$ \ldots$	&	$ \ldots$	&	$ \ldots$	\\
		&	-13.0	&	2.0	&	2.0	&	&	$ \ldots$	&	$ \ldots$	&	$ \ldots$ 	&	&	$ \ldots$	&	$ \ldots$	&	$ \ldots$	&	&	$ \ldots$	&	$ \ldots$	&	$ \ldots$	&	&	$ \ldots$	&	$ \ldots$	&	$ \ldots$	&	&	$ \ldots$	&	$ \ldots$	&	$ \ldots$	\\
		&	-9.8	&	12.2	&	2.2	&	&	$ \ldots$	&	$ \ldots$	&	$ \ldots$ 	&	&	-10.5	&	0.10	&	1.6	&	&	-9.1	&	51.9	&	2.5	&	&	$ \ldots$	&	$ \ldots$	&	$ \ldots$	&	&	$ \ldots$	&	$ \ldots$	&	$ \ldots$	\\
		&	-6.4	&	4.2	&	1.4	&	&	$ \ldots$	&	$ \ldots$	&	$ \ldots$ 	&	&	$ \ldots$	&	$ \ldots$	&	$ \ldots$	&	&	$ \ldots$	&	$ \ldots$	&	$ \ldots$	&	&	$ \ldots$	&	$ \ldots$	&	$ \ldots$	&	&	$ \ldots$	&	$ \ldots$	&	$ \ldots$	\\
		&	-3.6	&	4.3	&	2.5	&	&	$ \ldots$	&	$ \ldots$	&	$ \ldots$ 	&	&	$ \ldots$	&	$ \ldots$	&	$ \ldots$	&	&	-4.3	&	26.0	&	1.9	&	&	$ \ldots$	&	$ \ldots$	&	$ \ldots$	&	&	$ \ldots$	&	$ \ldots$	&	$ \ldots$	\\
		&	1.1	&	6.4	&	2.5	&	&	$ \ldots$	&	$ \ldots$	&	$ \ldots$ 	&	&	$ \ldots$	&	$ \ldots$	&	$ \ldots$	&	&	$ \ldots$	&	$ \ldots$	&	$ \ldots$	&	&	$ \ldots$	&	$ \ldots$	&	$ \ldots$	&	&	$ \ldots$	&	$ \ldots$	&	$ \ldots$	\\
		&	4.9	&	40.3	&	1.6	&	&	$ \ldots$	&	$ \ldots$	&	$ \ldots$ 	&	&	6.0	&	0.10	&	1.6	&	&	5.0	&	228.0	&	2.5	&	&	4.1	&	67.3	&	1.5	&	&	$ \ldots$	&	$ \ldots$	&	$ \ldots$	\\
\enddata
		\tablenotetext{a}{{\scriptsize The column density for CN includes \textit{J}=0 and \textit{J}=1 for detections. For upper limits, the column density for CN includes \textit{J}=0 (at the maximum depth at the largest CH component with \textit{b}=1 km s$^{-1}$) and includes the predicted \textit{J}=1 column density, 50\% of the \textit{J}=0 column density.}}
		\tablenotetext{b}{{\scriptsize No additional data.}}
	\end{deluxetable}
\end{longrotatetable}
\newpage
\begin{deluxetable}{lcc}
		\tabletypesize{\small }
	\tablecaption{GOT C+ Pointings \label{Herschellos}}
	\tablehead{\colhead{LOS-ID}  & \colhead{\textit{l} }  & \colhead{\textit{b}}   
	}
	\startdata
	G014.8-1.0	&	14.7826	&	-1.0	\\
	G010.4+0.0	&	10.4348	&	0.0	\\
	G010.4+0.0	&	10.4348	&	0.0	\\
	G015.7+1.0	&	15.6522	&	1.0	\\
	G020.0+0.0	&	20.0000	&	0.0	\\
	G032.6+0.0	&	32.5532	&	0.0	\\
	G091.7+1.0	&	91.6981	&	1.0	\\
	G109.8+0.0	&	109.8113	&	0.0	\\
	G207.2-1.0	&	207.1700	&	-1.0	\\
	G225.3+0.0	&	225.2830	&	0.0	\\
	\enddata
\end{deluxetable}
\clearpage
\pagebreak
\startlongtable
\begin{deluxetable}{lccccccccccccccccccccc}
\tabletypesize{\scriptsize }
\tablecaption{Velocities of Components with Corresponding Emission Features in the Inner Galaxy \label{outertab}}
\tablehead{\multicolumn{2}{l}{Species} &  \multicolumn{16}{c}{$V_{lsr}$ (km s$^{-1}$)}&&}	
	\startdata
	\\
	\multicolumn{20}{c}{\underline{G010.4+0.0}}	\\%
	C$^{+}$	&	-38.9	&	$\ldots$	&	$\ldots$	&	$\ldots$	&	$\ldots$	&	5.8	&	$\ldots$	&	$\ldots$	&$\ldots$	&	25.0		&	$\ldots$	&	45.3	&	65.0		&	$\ldots$&	73.7	&	$\ldots$	&	$\ldots$	&	$\ldots$	&	$\ldots$	\\
	\ion{H}{1} 	&	$\ldots$	&	-25.8	&	-19.5	&	-10.0		&	1.2	&	$\ldots$	&	$\ldots$	&	11.8	&	20.9		&$\ldots$	&	35.4	&	45.3	&	65.7	&	$\ldots$&		75.1	&	90.7	&	122.4	&	$\ldots$	&	150.9	\\
	$^{12}$CO	&	$\ldots$	&	$\ldots$	&	$\ldots$	&	-11.0		&	$\ldots$	&	$\ldots$	&	6.8	&	$\ldots$	&	20.4		& 26.7	&	34.0	&		$\ldots$	&	$\ldots$&	69.5		&	$\ldots$&	$\ldots$	&	122.9	&	145.7	&	153.2	\\
	$^{13}$CO	&	$\ldots$	&	$\ldots$	&	$\ldots$	&	-11.0		&	$\ldots$	&	$\ldots$	&	6.3	&	$\ldots$	&	20.5 & 23.4		&	34.3	&		$\ldots$	&	$\ldots$&	69.0	&	$\ldots$	&	$\ldots$	&	$\ldots$	&	$\ldots$	&	$\ldots$	\\
\enddata
\tablenotetext{a}{Occasionally two nearby components need to be combined to create agreement with the features in another species. This is done by weighting the $V_{lsr}$ of the components by their relative column densities and taking an average.}	
\end{deluxetable}
\newpage
\startlongtable
\begin{deluxetable}{lccccccccccccccccccc}
\tabletypesize{\scriptsize}
\tablecaption{Velocities of Components with Corresponding Emission Features in the Outer Galaxy \label{innertab}}
\tablehead{Species & & \multicolumn{7}{c}{$V_{lsr}$ (km s$^{-1}$)}&&&&&&&&&&&}
\startdata
\\		\multicolumn{20}{c}{\underline{G091.7+1.0}}	\\	%
	C$^{+}$	&	$\ldots$	& -100.7 &	$\ldots$	&	$\ldots$	&	$\ldots$	&	$\ldots$	&	$\ldots$	&	$\ldots$	&	$\ldots$	&	$\ldots$	&	$\ldots$	&\\
																									\ion{H}{1} 	&	-111.9	&	$\ldots$&	-93.9	&	-82.2	&	-70.5	&	-54.0	&	-43.5	&	-32.6	&	-10.1	&	0.1	&	5.0		&\\
																									\\
																									\multicolumn{20}{c}{\underline{BD+49$^{\circ}$ 3482}}	\\	
																									\ion{Ca}{2}	&	$\ldots$&	$\ldots$	&	$\ldots$	&	$\ldots$	&	$\ldots$	&	$\ldots$	&	$\ldots$	&	$\ldots$	&	$\ldots$	&	1.6	&	5.4\tablenotemark{a}		&\\
																									\ion{K}{1}	&	$\ldots$&	$\ldots$	&	$\ldots$	&	$\ldots$	&	$\ldots$	&	$\ldots$	&	$\ldots$	&	$\ldots$	&	$\ldots$	&	1.2	&	$\ldots$	&\\
\enddata
																									\tablenotetext{a}{Occasionally two nearby components need to be combined to create agreement with the features in another species. This is done by weighting the $V_{lsr}$ of the components by their relative column densities and taking an average.}

\tablenotetext{b}{This component is likely associated with the emission component through additional correspondence with literature data. (See text for further details.)}
\end{deluxetable}
\newpage

\begin{deluxetable}{lcccccccccccc}
\tabletypesize{\scriptsize}
\tablecolumns{12} 
\tablewidth{0pt} 
\tablecaption{Comparison of Component Structure: G010.4+0.0 \label{G010.4+0.0table}}
\tablehead{
&	\colhead{G010.4+0.0  }	&	\colhead{HD 165783}	&	\colhead{HD 165918}	&			\\
\colhead{$V_{lsr}$}	&	\colhead{GOT C+}	&	\colhead{McDonald}	&	\colhead{McDonald}	&	\colhead{Comments}		\\
\colhead{(km s$^{-1}$)}	&	\colhead{Components}	&	\colhead{Components}	&	\colhead{Components}	&					
}

\startdata
-20	&	\ion{H}{1}	&	\ion{Ca}{2}	&	\ion{Ca}{2}	&	Warm neutral		\\
-10	&	\ion{H}{1}, $^{12}$CO	&	\ion{Ca}{2}, CH$^{+}$	&	\ion{Ca}{2}, CH$^{+}$	&	Molecular cloud edge		\\
1.0	&	\ion{H}{1}	&	\ion{Ca}{2}	&	\ion{Ca}{2}	&	Warm neutral		\\
5.0	&	C$^{+}$	&	\ion{Ca}{2}, CH$^{+}\!,$ CH	&	\ion{Ca}{2}, CH$^{+}\!,$ CH	&	Cold neutral, CO-dark		\\
7.0	&	$^{12}$CO	&	\ion{Ca}{2}, CH$^{+}\!,$ CH	&	$\ldots$	&	Molecular cloud edge		\\
10	&	\ion{H}{1}	&	\ion{Ca}{2}	&	\ion{Ca}{2}	&	Warm neutral		\\
25	&	C$^{+}\!,$ \ion{H}{1},  $^{12}$CO	&	\ion{Ca}{2}	&	$\ldots$	&	Molecular cloud edge		\\
35	&	\ion{H}{1}, $^{12}$CO	&	\ion{Ca}{2}	&	$\ldots$	&	Molecular cloud edge		\\
45	&	C$^{+}\!,$ \ion{H}{1}	&	\ion{Ca}{2}	&	$\ldots$	&	Cold neutral		\\
\enddata
\end{deluxetable}

\begin{deluxetable}{lcccccccccccc}
	\tablecolumns{12} 
	\tablewidth{0pt} 
	\tablecaption{Comparison of Component Structure: G014.8-1.0 \label{G014.8table}}
	\tablehead{
		&	\colhead{G014.8-1.0}	&	\colhead{HD 168607}	&					\\
		\colhead{$V_{lsr}$}	&	\colhead{GOT C+}	&	\colhead{McDonald}	&	\colhead{Comments}				\\
		\colhead{(km s$^{-1}$)}	&	\colhead{Components}	&	\colhead{Components}	&				
	}
	
	\startdata
	-24	&	\ion{H}{1}, $^{12}$CO	&	\ion{Ca}{2}	&	Molecular cloud edge				\\
	-9.0	&	\ion{H}{1}	&	\ion{Ca}{2}	&	Warm neutral				\\
	6.0	&	\ion{H}{1}	&	\ion{Ca}{2}, CH$^{+}\!,$ CH	&	Cold neutral, CO-dark				\\
	15	&	\ion{H}{1}	&	\ion{Ca}{2}, CH$^{+}\!,$ CH	&	Cold neutral, CO-dark				\\
	19	&	C$^{+}\!,$ $^{12}$CO, $^{13}$CO, C$^{18}$O	&	\ion{Ca}{2}, CH, CN	&	Molecular cloud				\\
	23	&	\ion{H}{1}	&	\ion{Ca}{2}, CH$^{+}\!,$ CH	&	Cold neutral, CO-dark				\\
	30	&	$^{12}$CO, $^{13}$CO, C$^{18}$O	&	\ion{Ca}{2}	&	Molecular cloud				\\
	37	&	\ion{H}{1}	&	\ion{Ca}{2}	&	Warm neutral				\\
	46	&	$^{12}$CO	&	\ion{Ca}{2}	&	Molecular cloud edge				\\
	53	&	\ion{H}{1}, $^{12}$CO	&	\ion{Ca}{2}	&	Molecular cloud edge				\\
	\enddata
\end{deluxetable}

\clearpage

\begin{deluxetable}{ccccccc}
		\tabletypesize{\tiny}
		\tablecolumns{8} 
		\tablewidth{0pt} 
		\tablecaption{Summary\tablenotemark{a} \label{Sumres7}}
		\tablehead{
\colhead{Type of Gas}&\colhead{ }& \colhead{Diffuse Atomic}& \multicolumn2c{Diffuse Molecular }&\colhead{ }&\colhead{ }\\
\colhead{ }&\colhead{ }& \colhead{ }& \colhead{\textit{n}$\sim$100 cm$^{-3}$}&\colhead{\textit{n}$\sim$300 cm$^{-3}$}&\colhead{Total }&\colhead{Additional}\\
			&\colhead{Species}	&	\colhead{\ion{Ca}{2} } &	\colhead{\ion{K}{1} or CH or CH$^{+}$ }&	\colhead{CN and }  &\colhead{Associated}&  \colhead{Unassociated} \\
			\colhead{ }&\colhead{Detected}&  \colhead{only }		& \colhead{and \ion{Ca}{2}  }	& \colhead{\ion{Ca}{2} or \ion{K}{1} or CH or CH$^{+}$ }	&\colhead{  Components } &	\colhead{Emission Components}
		}
		\startdata
Diffuse Atomic (WNM)	&	\ion{H}{1} only	&	7	&	12(5\tablenotemark{b})	&	1	&	20	&	 31 (12\tablenotemark{b}) \\
&		&		&		&		&		&	\\
Diffuse Atomic (CNM)	&	C$^{+}$ only	&	1	&	1	&	0	&	2	&	5 (1\tablenotemark{c}) \\
&	C$^{+}$ +\ion{H}{1}	&	1	&	1	&	0	&	2	&	 6 (1\tablenotemark{b}) \\
&		&		&		&		&		&	\\
&		&		&		&		&		&	\\
Molecular gas	&	$^{12}$CO only	&	1	&	0	&	1	&	2	&	 3  \\
&	C$^{+}$ + $^{12}$CO	&	0	&	1(1\tablenotemark{c})	&	0	&	1	&	0\\
\enddata
		\tablenotetext{a}{This table specifies the number of emission components that have been associated with at least one absorption component near the same $V_{lsr}$.}
		\tablenotetext{b}{Number of included components without CO data.}
		\tablenotetext{c}{Number of included components without H I data.}
		\tablenotetext{d}{Number of included components that are also seen in a second nearby sight line.}
\end{deluxetable}
	\newpage

\begin{figure}
	\figurenum{1}
	\plotone{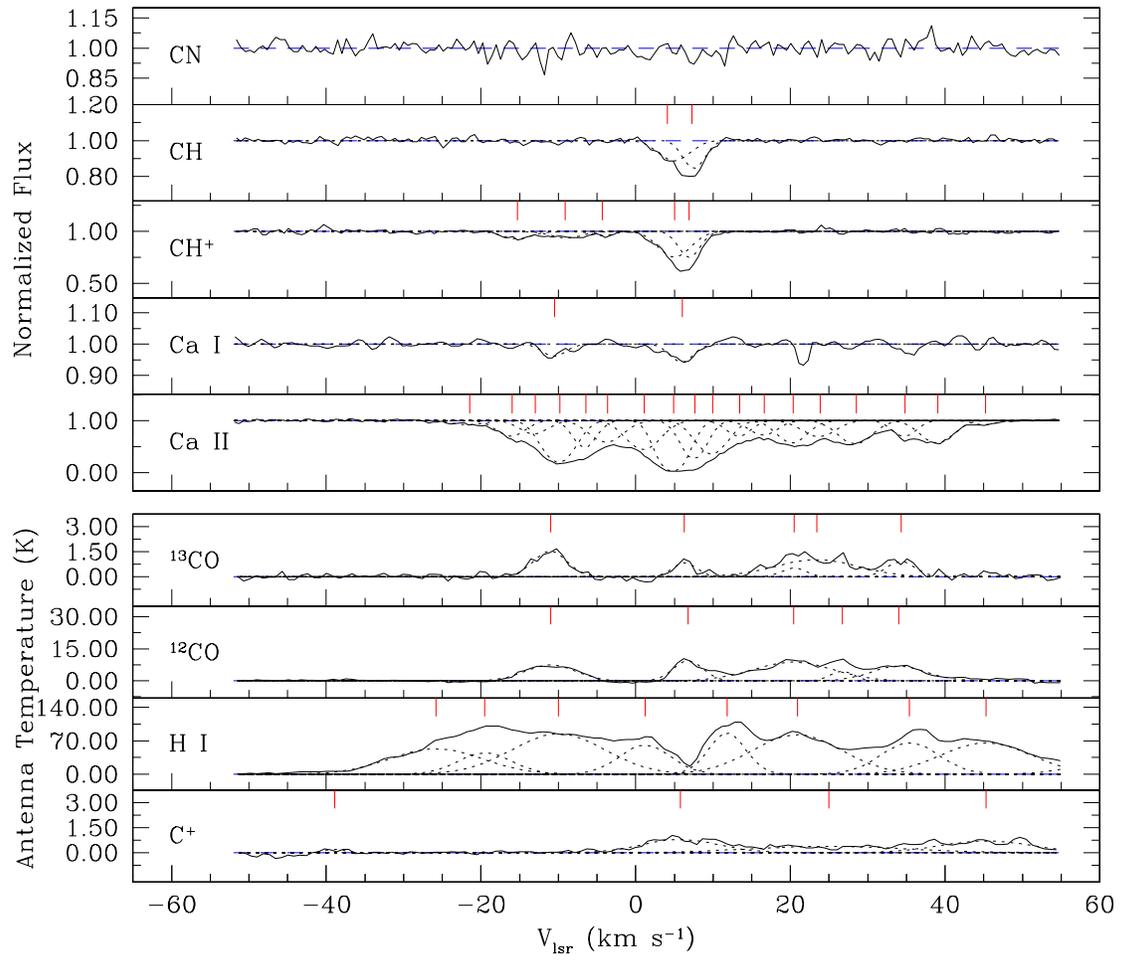}
	\caption{(Top) Absorption spectra in $V_{lsr}$ toward HD 165783. (Bottom) Emission spectra in $V_{lsr}$ from the closest pointing, G010.4+0.0. The individual components are shown as dotted lines in each species and indicated by red ticks above the fit.}
\end{figure} 

\begin{figure}
	\figurenum{2}
	\plotone{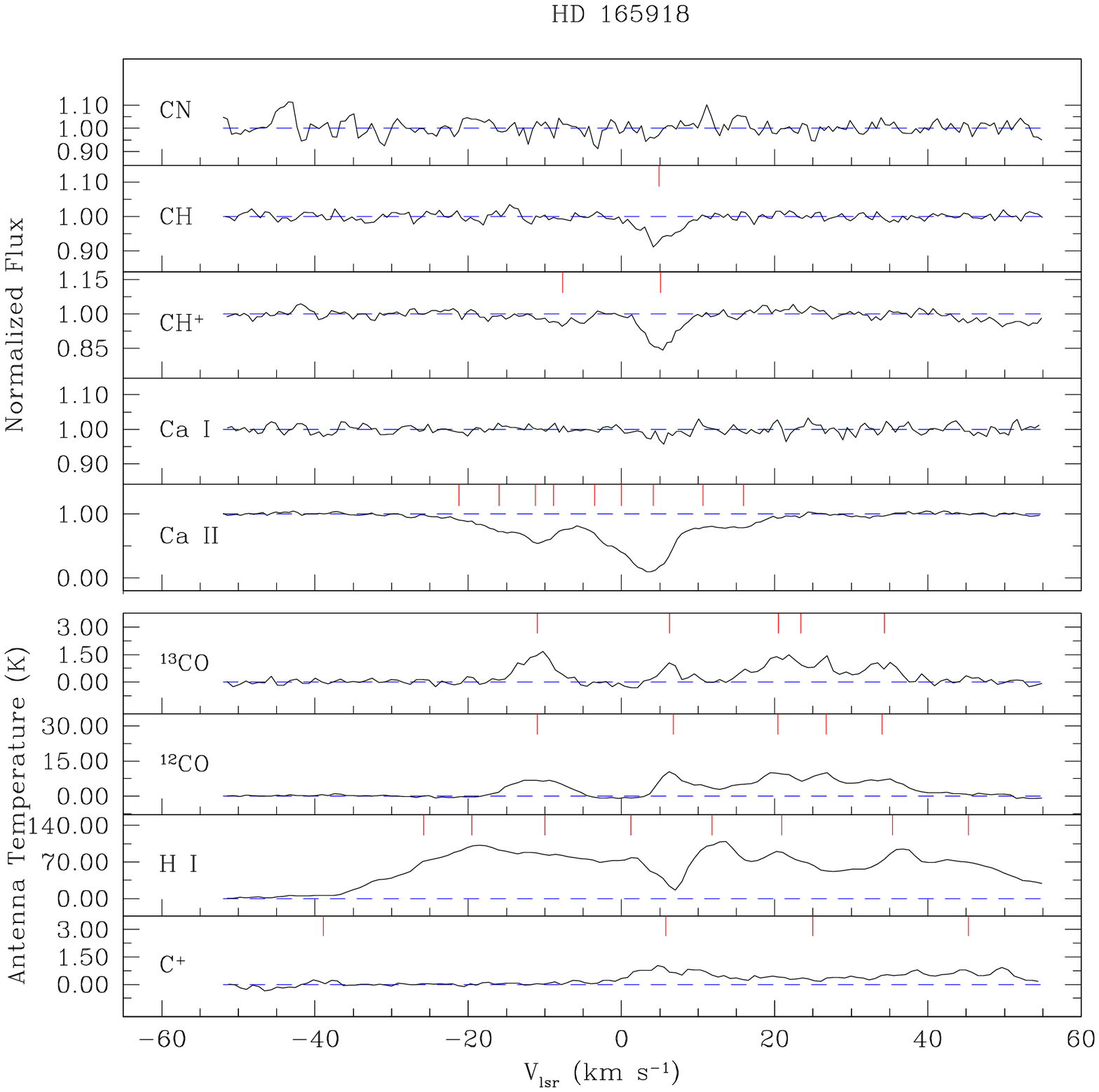}
	\caption{(Top) Absorption spectra in $V_{lsr}$ toward HD 165918. (Bottom) Emission spectra in $V_{lsr}$ from the closest pointing, G010.4+0.0. The red ticks indicate the locations of fit components. }
\end{figure} 
 
 \begin{figure}
 	\figurenum{3}
 	\plotone{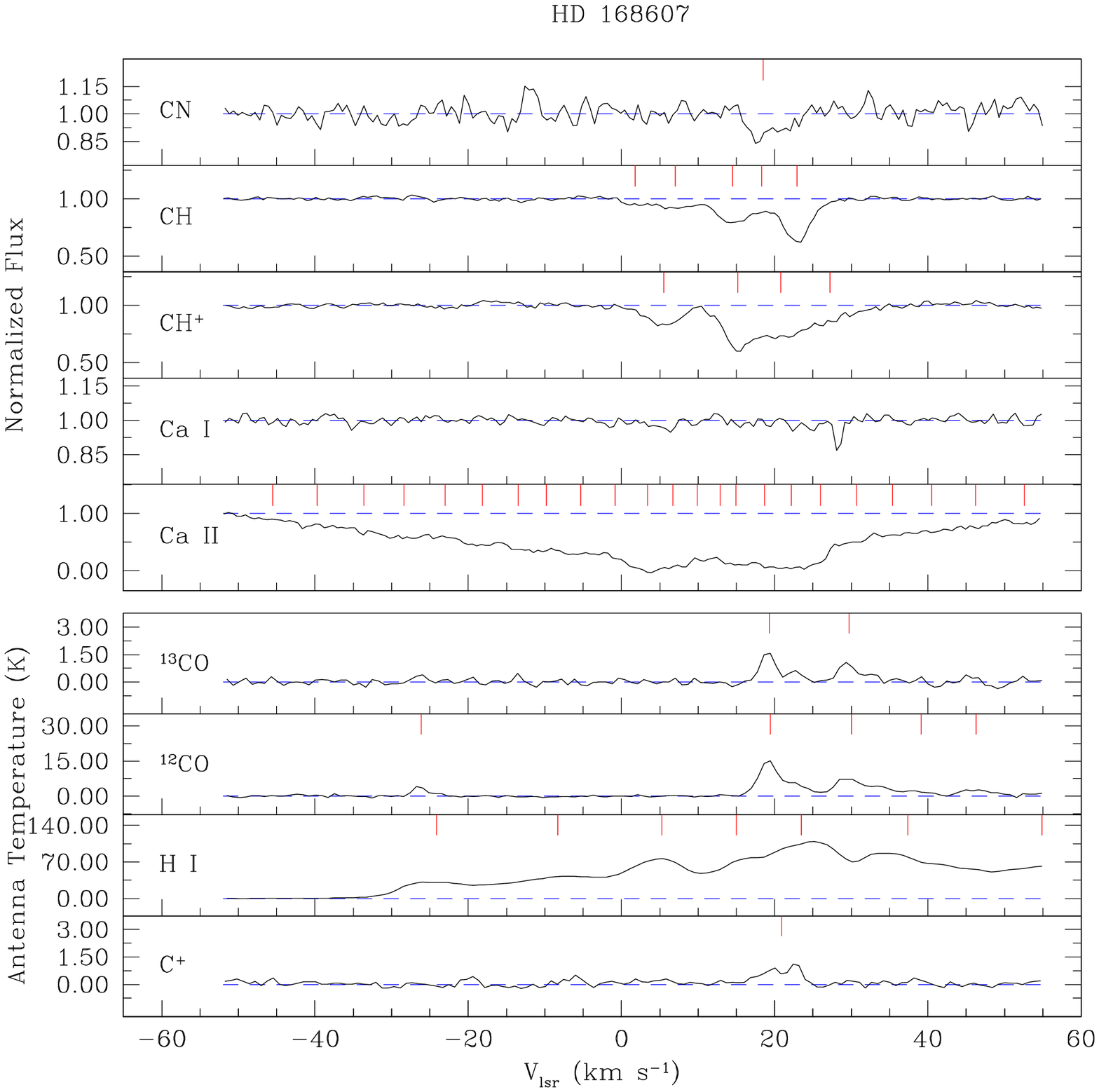}
 	\caption{(Top) Absorption spectra in $V_{lsr}$ toward HD 168607. (Bottom) Emission spectra in $V_{lsr}$ from the closest pointing, G014.8-1.0. The red ticks indicate the locations of fit components.  }
 \end{figure} 

\begin{figure}
	\figurenum{4}
	\plotone{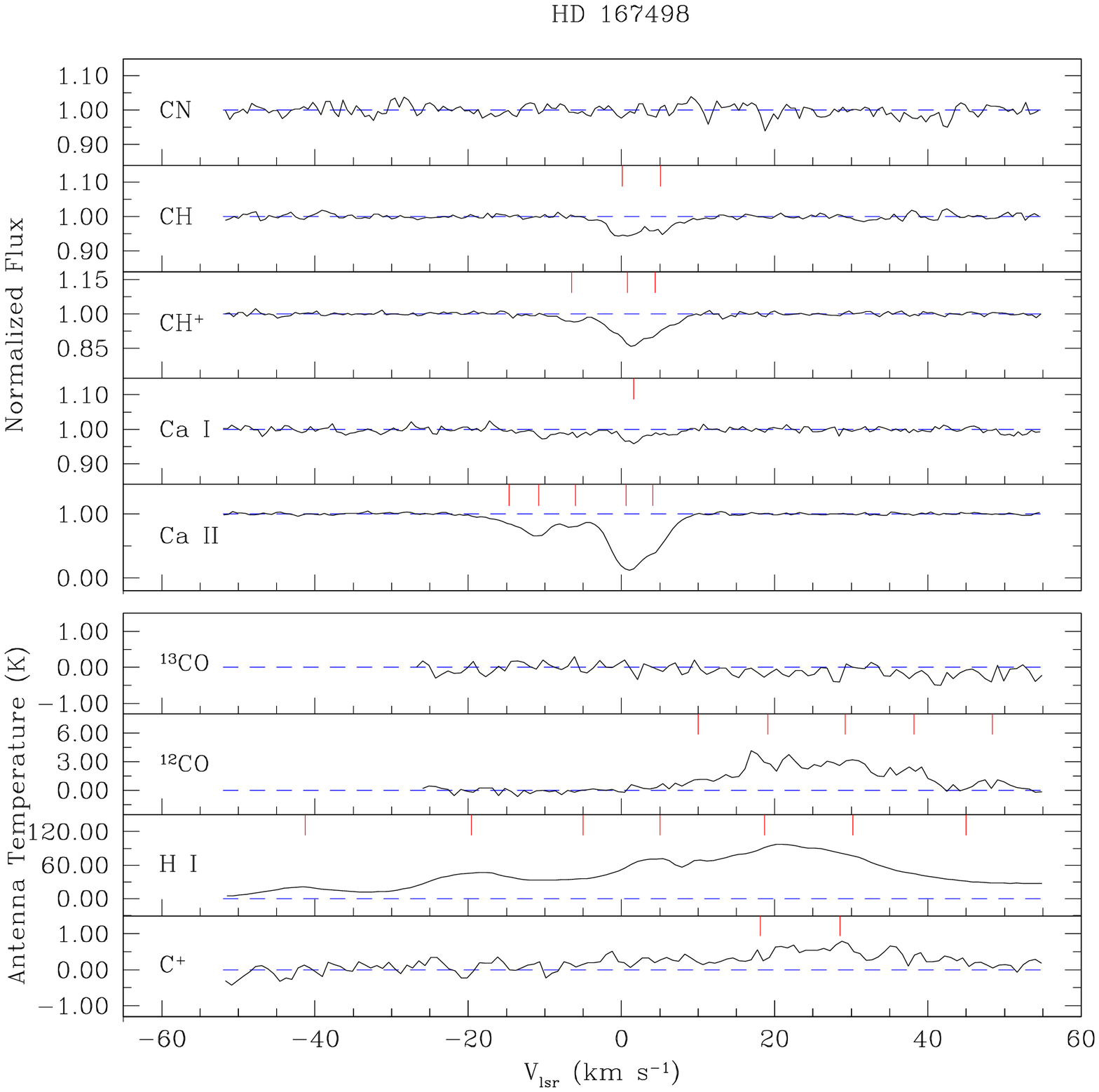}
	\caption{(Top) Absorption spectra in $V_{lsr}$ toward HD 167498. (Bottom) Emission spectra in $V_{lsr}$ from the closest pointing, G015.7+1.0. The red ticks indicate the locations of fit components.  }
\end{figure} 

\begin{figure}
	\figurenum{5}
	\plotone{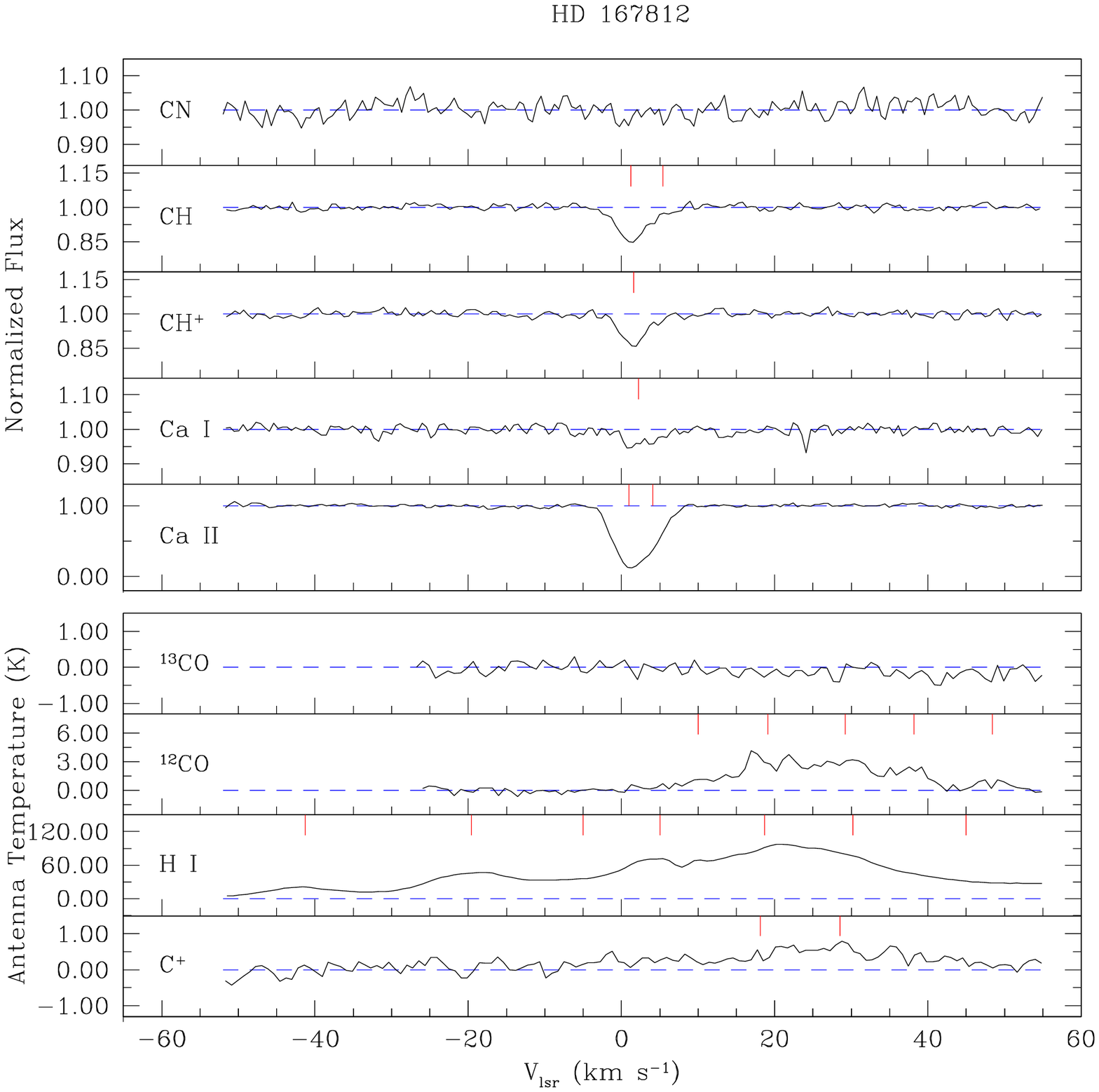}
	\caption{(Top) Absorption spectra in $V_{lsr}$ toward HD 167812. (Bottom) Emission spectra in $V_{lsr}$ from the closest pointing, G015.7+1.0. The red ticks indicate the locations of fit components. }
\end{figure} 
\begin{figure}
	\figurenum{6}
	\plotone{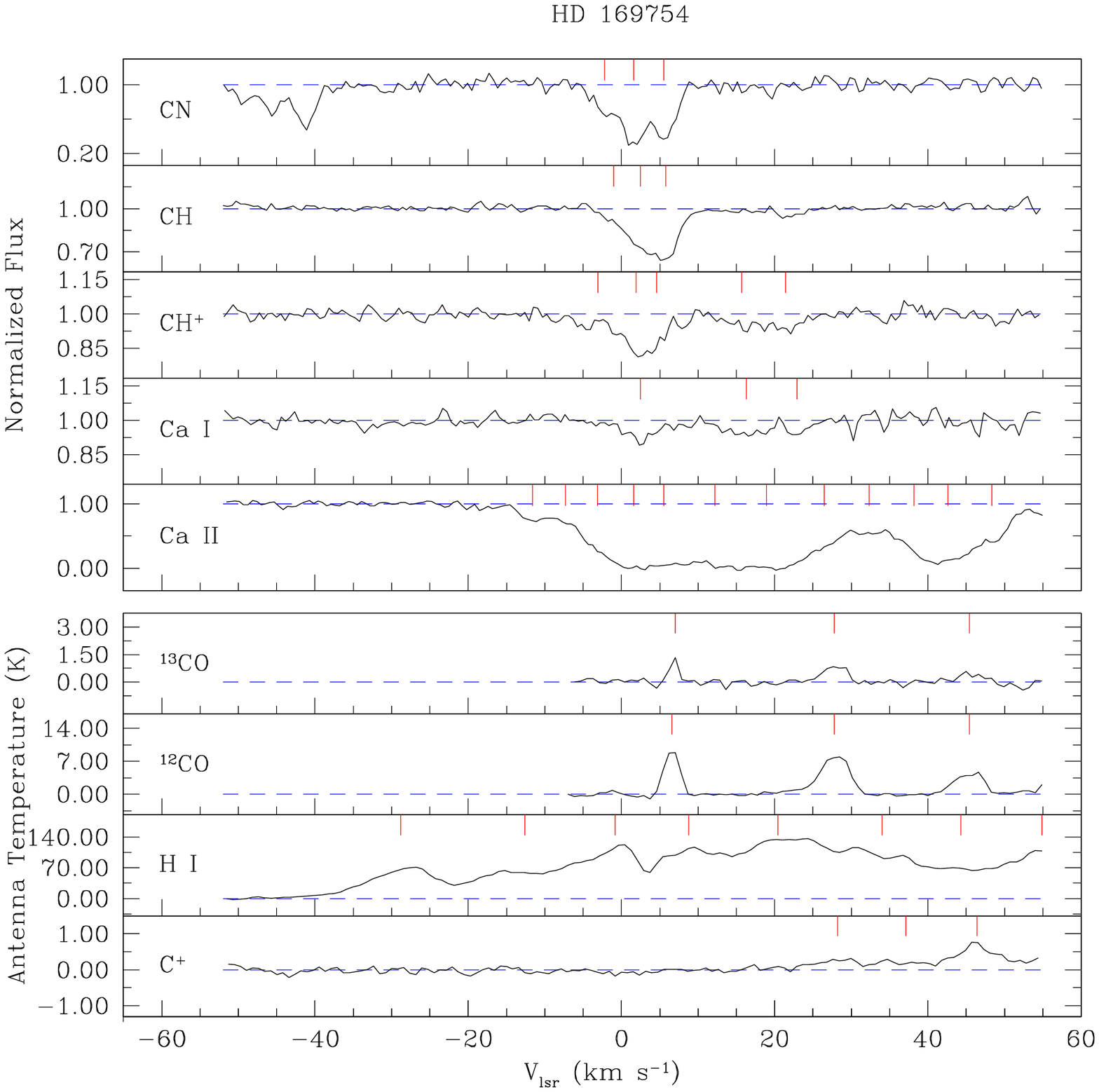}
	\caption{(Top) Absorption spectra in $V_{lsr}$ toward HD 169754. (Bottom) Emission spectra in $V_{lsr}$ from the closest pointing, G020.0+0.0.The red ticks indicate the locations of fit components. }
\end{figure} 
\begin{figure}
	\figurenum{7}
	\plotone{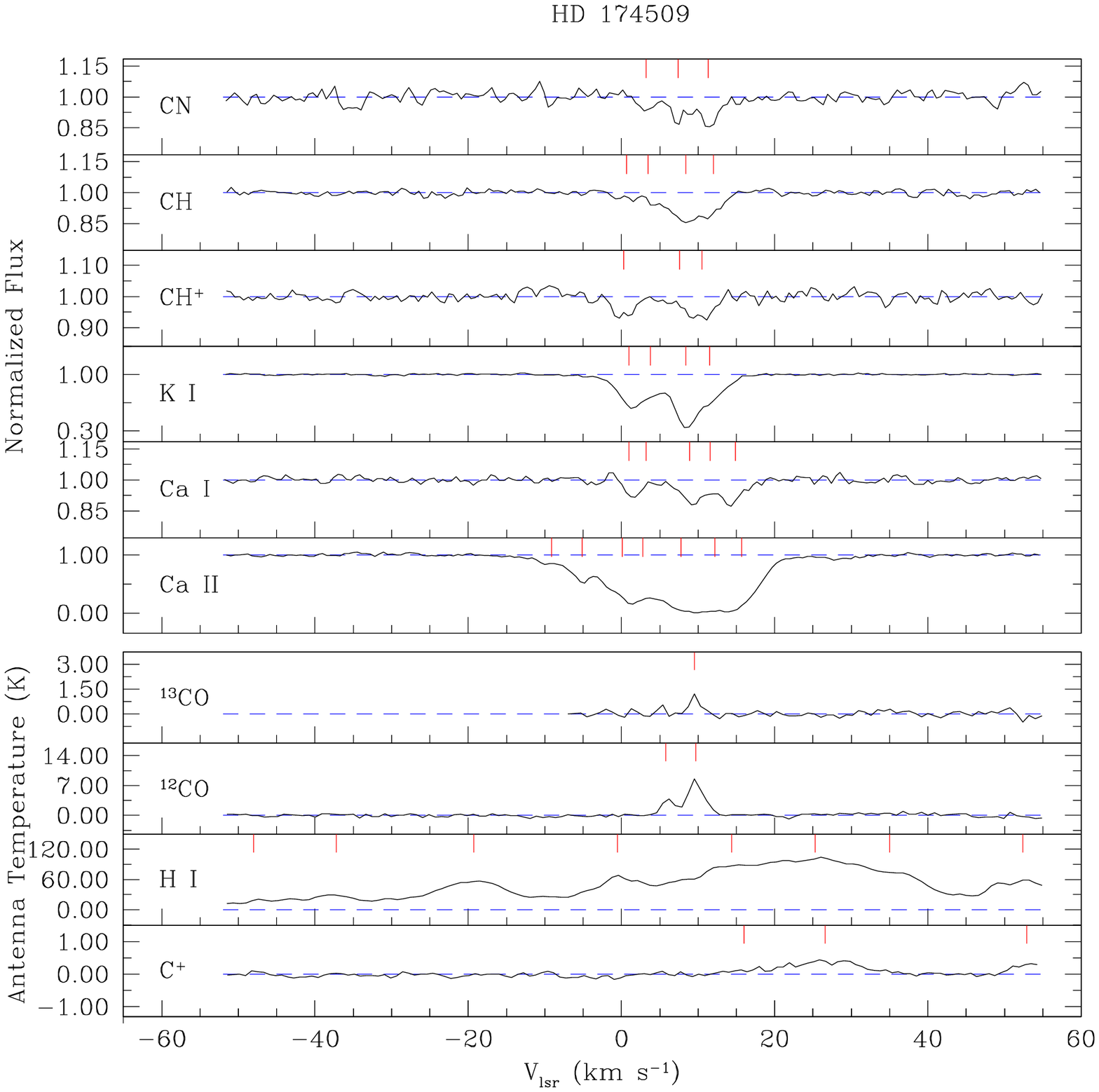}
	\caption{(Top) Absorption spectra in $V_{lsr}$ toward HD 174509. (Bottom) Emission spectra in $V_{lsr}$ from the closest pointing, G032.6+0.0. The red ticks indicate the locations of fit components.}
\end{figure} 
\begin{figure}
	\figurenum{8}
	\plotone{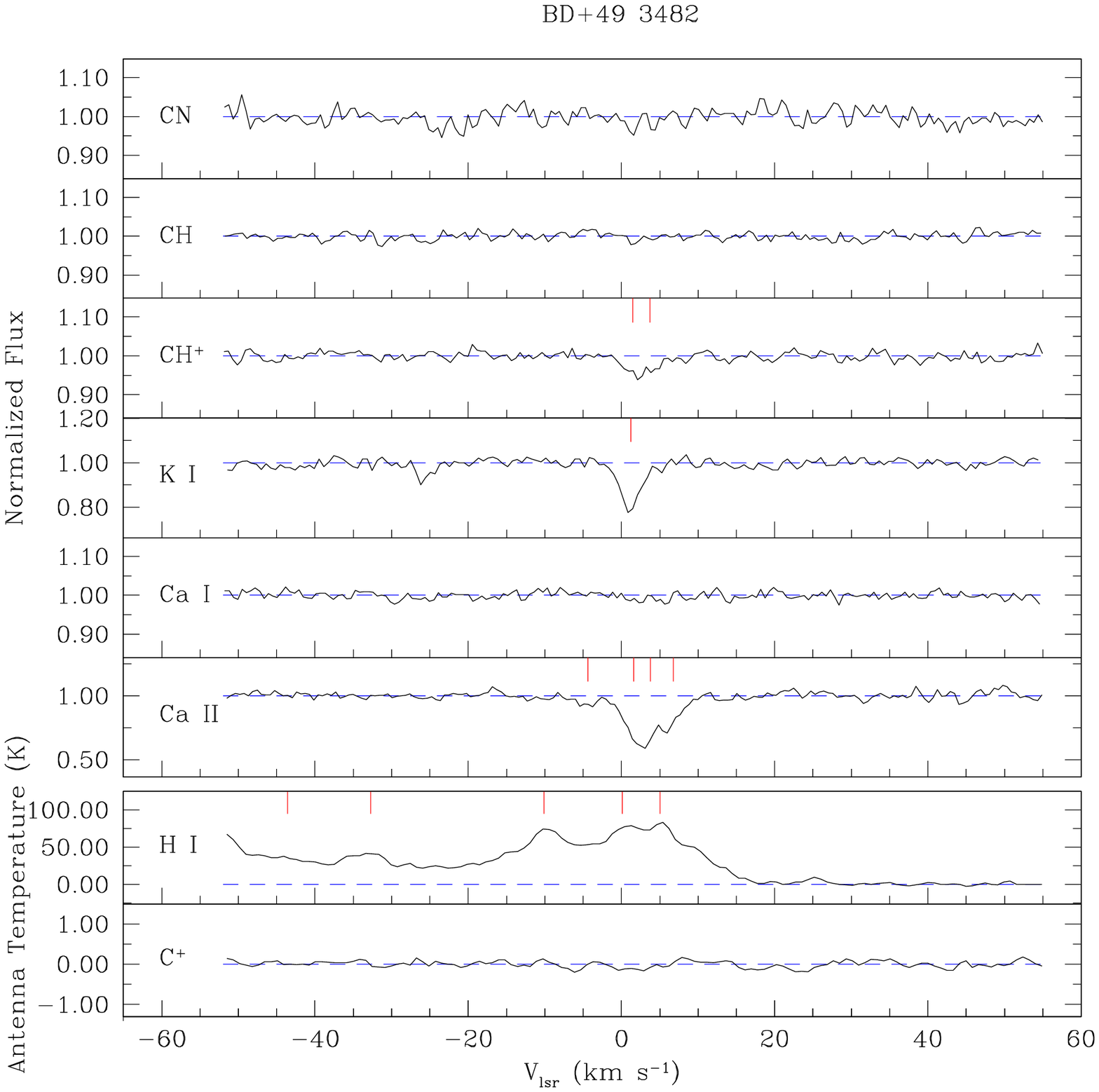}
	\caption{(Top) Absorption spectra in $V_{lsr}$ toward BD+49 3482. (Bottom) Emission spectra in $V_{lsr}$ from the closest pointing, G091.7+1.0. The red ticks indicate the locations of fit components.}
\end{figure} 
\begin{figure}
	\figurenum{9}
	\plotone{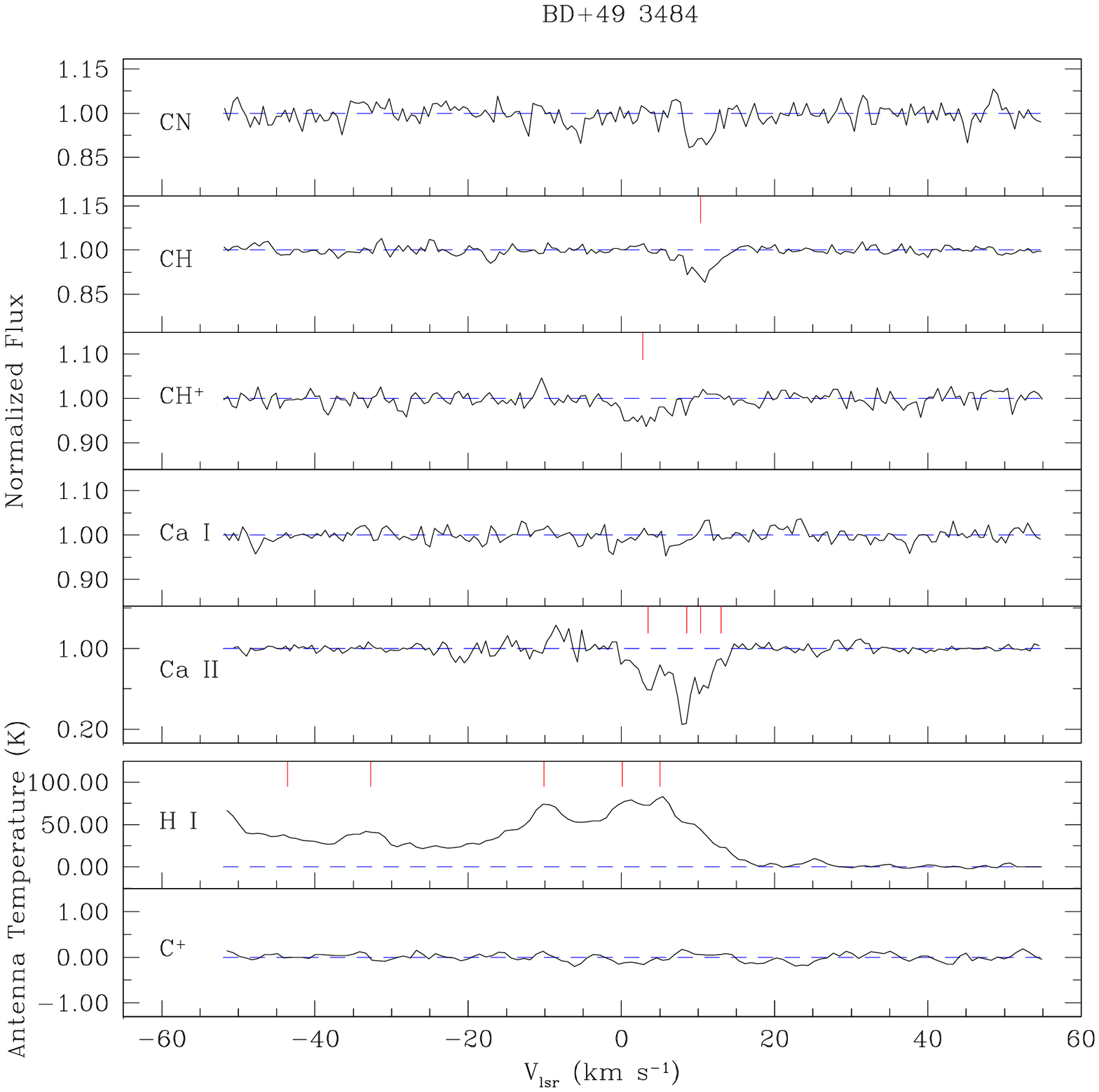}
	\caption{(Top) Absorption spectra in $V_{lsr}$ toward BD+49 3484. (Bottom) Emission spectra in $V_{lsr}$ from the closest pointing, G091.7+1.0. The red ticks indicate the locations of fit components.}
\end{figure} 
\begin{figure}
	\figurenum{10}
	\plotone{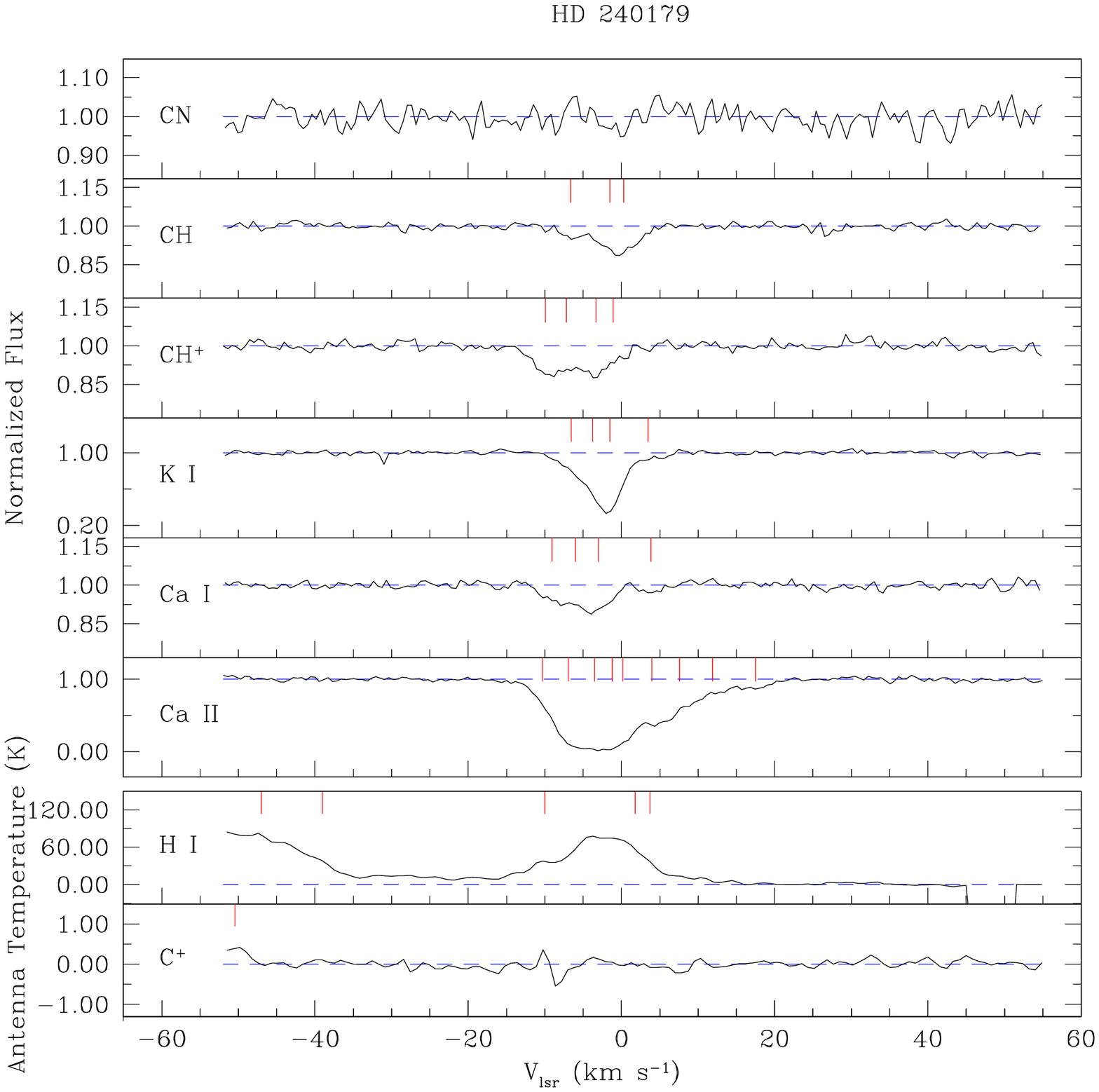}
	\caption{(Top) Absorption spectra in $V_{lsr}$ toward HD 240179. (Bottom) Emission spectra in $V_{lsr}$ from the closest pointing, G109.8+0.0. The red ticks indicate the locations of fit components.}
\end{figure} 
\begin{figure}
	\figurenum{11}
	\plotone{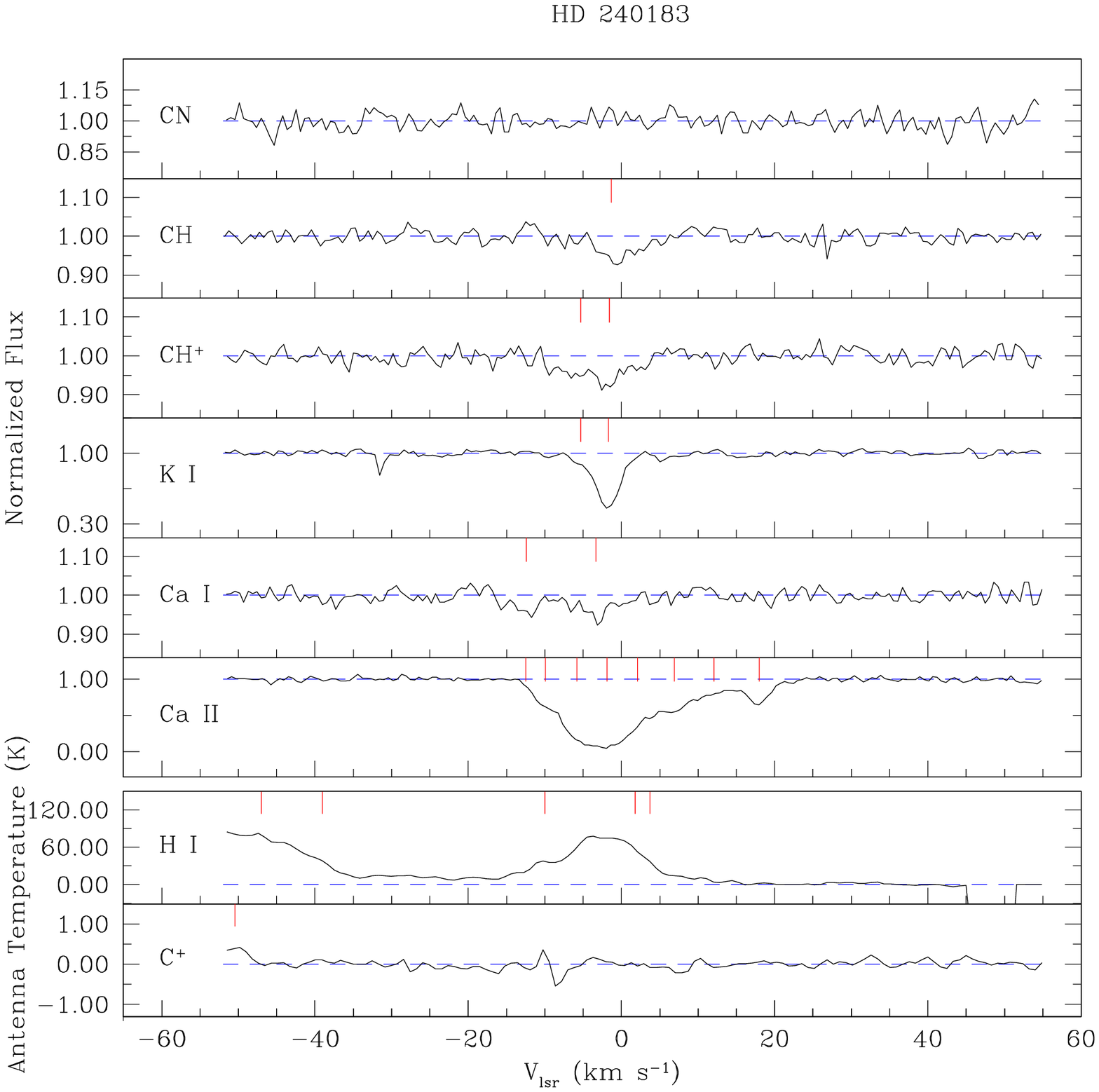}
	\caption{(Top) Absorption spectra in $V_{lsr}$ toward HD 240183. (Bottom) Emission spectra in $V_{lsr}$ from the closest pointing, G109.8+0.0. The red ticks indicate the locations of fit components.}
\end{figure} 
\begin{figure}
	\figurenum{12}
	\plotone{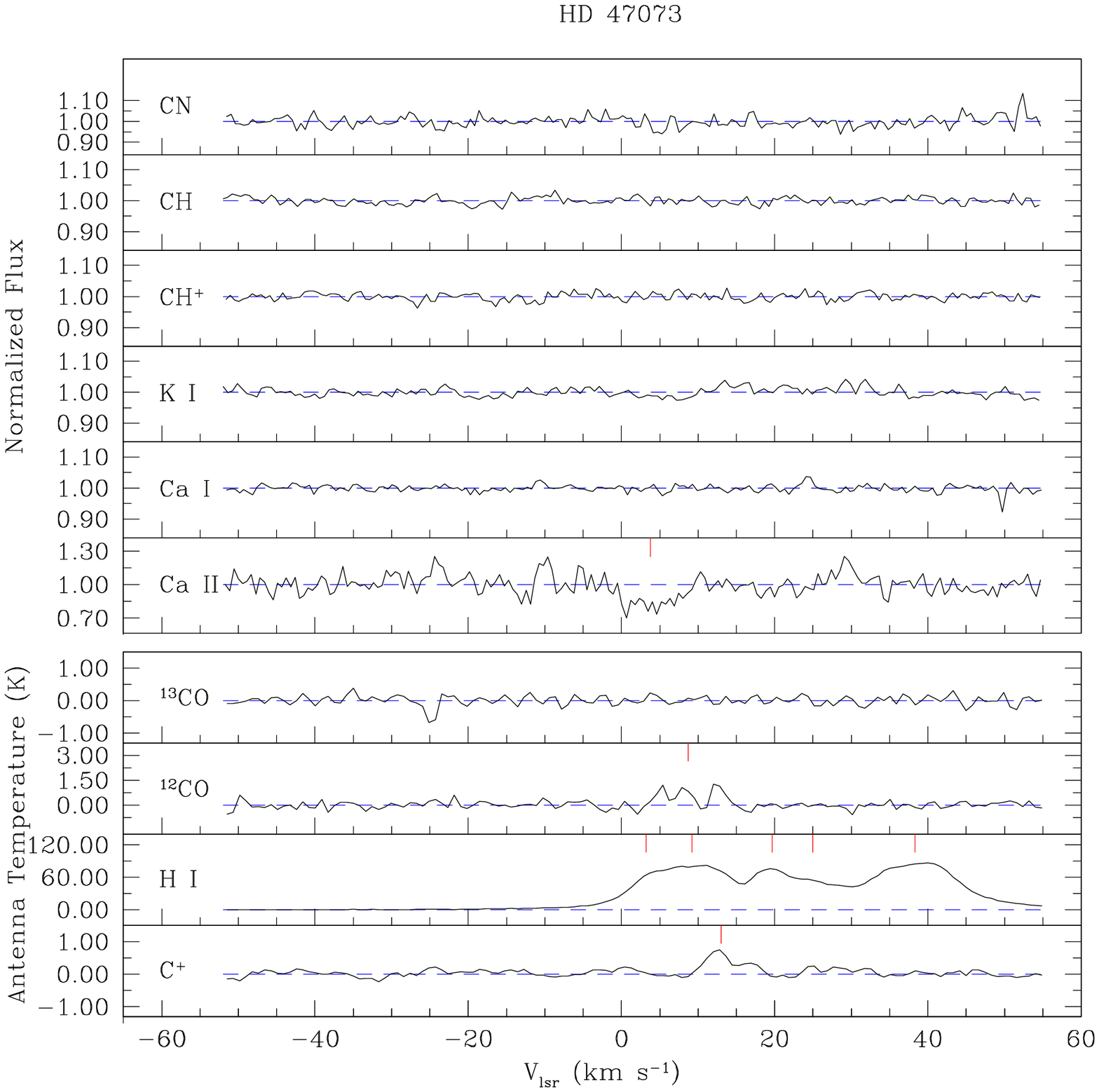}
	\caption{(Top) Absorption spectra in $V_{lsr}$ toward HD 47073. (Bottom) Emission spectra in $V_{lsr}$ from the closest pointing, G207.2-1.0. The red ticks indicate the locations of fit components.}
\end{figure} 
\begin{figure}
	\figurenum{13}
	\plotone{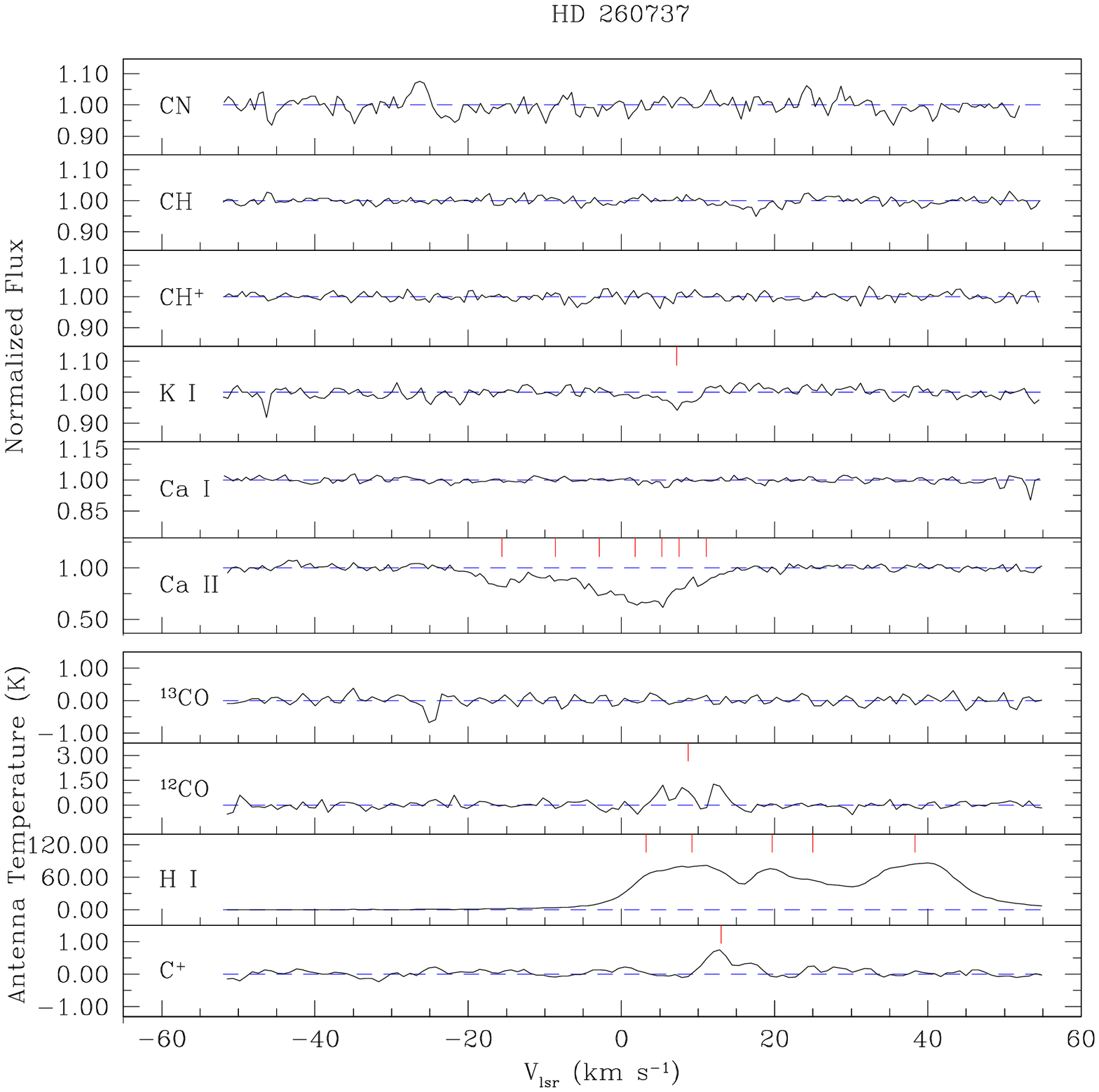}
	\caption{(Top) Absorption spectra in $V_{lsr}$ toward HD 260737. (Bottom) Emission spectra in $V_{lsr}$ from the closest pointing, G207.2-1.0. The red ticks indicate the locations of fit components.}
\end{figure} 
\begin{figure}
	\figurenum{14}
	\plotone{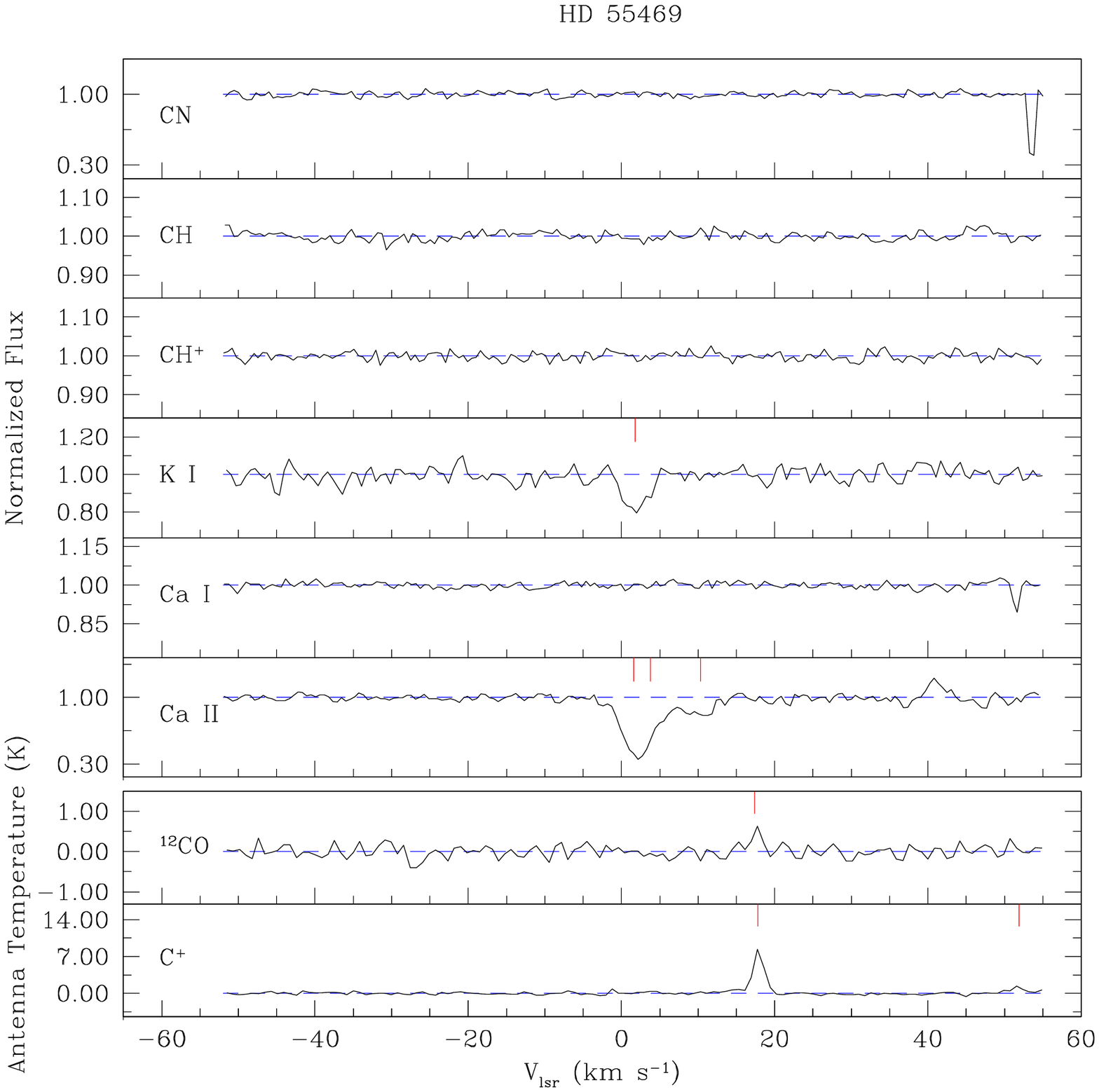}
	\caption{(Top) Absorption spectra in $V_{lsr}$ toward HD 55469. (Bottom) Emission spectra in $V_{lsr}$ from the closest pointing, G225.3+0.0. The red ticks indicate the locations of fit components.}
\end{figure} 
\begin{figure}
	\figurenum{15}
	\plotone{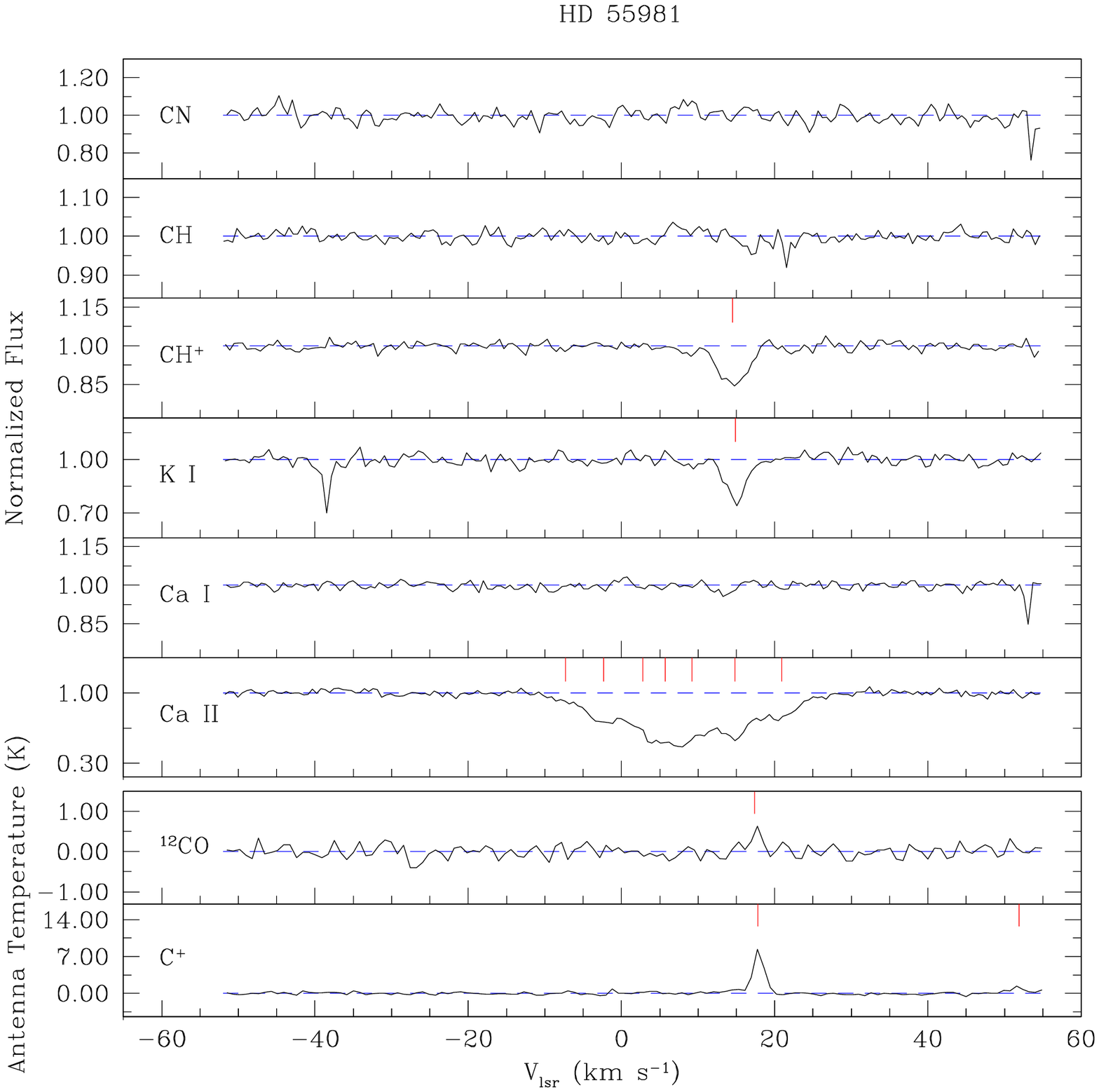}
	\caption{(Top) Absorption spectra in $V_{lsr}$ toward HD 55981. (Bottom) Emission spectra in $V_{lsr}$ from the closest pointing, G225.3+0.0. The red ticks indicate the locations of fit components.}
\end{figure} 

 \newpage 	\clearpage

\nocite{Tull1995}
\nocite{Wenger2000}
\nocite{Zhang2005}
\nocite{Schlingman2011}
\nocite{Dunham2011}
\nocite{Sheffer2008}
\nocite{Pan2005}
\nocite{Sheffer2008}
\nocite{Federman1994}
\nocite{Dobashi1994}
\nocite{Dobashi2005}
\nocite{Dobashi2011}
\nocite{Kim2004}
\nocite{Wouterloot1988}
\nocite{Wouterloot1989}
\nocite{Langer2010}
\nocite{2014A&A...561A.122L}
\nocite{2013A&A...554A.103P}
\nocite{2001ApJS..133..345W}
\nocite{Welty2003}
\nocite{Lynds1962}
\nocite{Stil2006}
\nocite{Pilbratt2010}
\nocite{Graauw2010}
\nocite{1982IAUS...98..261J}
\nocite{1988mcts.book.....H}
\nocite{2000A&A...355L..27H}
\nocite{2002yCat.2237....0D}
\nocite{2000PASP..112...50W}
\nocite{1955ApJS....2...41M}
\nocite{1999MSS...C05....0H}
\nocite{1993yCat.3135....0C}
\nocite{1997ESASP1200.....E}
\nocite{1968PASP...80..197G}
\nocite{2010AN....331..349H}
\nocite{1974AJ.....79.1022C}

\nocite{1994MNRAS.266..903C}
\nocite{1985ApJ...288..604C}
\nocite{1987ApJ...316L..71C}
\nocite{1982ApJ...260..124F}
\nocite{1984ApJ...283..626F}
\nocite{1994A&A...285..300G}
\nocite{1992A&A...257..245G}
\nocite{2009ApJ...696.1533H}
\nocite{1976ApJ...209..782K}
\nocite{1990ApJ...359L..19L}
\nocite{1979ApJ...233L.147L}
\nocite{1986A&A...160..157M}
\nocite{2003A&A...408..545W}
\nocite{1981ApJ...247..116W}

\nocite{2010ApJ...715.1370G}
\nocite{2011ApJ...743..174I}
\bibliographystyle{apj}
\bibliography{references}

\begin{thebibliography}{}
\expandafter\ifx\csname natexlab\endcsname\relax\def\natexlab#1{#1}\fi

\bibitem[{{Cannon} \& {Pickering}(1993)}]{1993yCat.3135....0C}
{Cannon}, A.~J., \& {Pickering}, E.~C. 1993, VizieR Online Data Catalog, 3135

\bibitem[{{Claria}(1974)}]{1974AJ.....79.1022C}
{Claria}, J.~J. 1974, \aj, 79, 1022

\bibitem[{{Crawford} {et~al.}(1994){Crawford}, {Barlow}, {Diego}, \&
  {Spyromilio}}]{1994MNRAS.266..903C}
{Crawford}, I.~A., {Barlow}, M.~J., {Diego}, F., \& {Spyromilio}, J. 1994,
  \mnras, 266, 903

\bibitem[{{Crutcher}(1985)}]{1985ApJ...288..604C}
{Crutcher}, R.~M. 1985, \apj, 288, 604

\bibitem[{{Crutcher} \& {Federman}(1987)}]{1987ApJ...316L..71C}
{Crutcher}, R.~M., \& {Federman}, S.~R. 1987, \apjl, 316, L71

\bibitem[{{de Graauw} {et~al.}(2010){de Graauw}, {Helmich}, {Phillips},
  {Stutzki}, {Caux}, {Whyborn}, {Dieleman}, {Roelfsema}, {Aarts}, {Assendorp},
  {Bachiller}, {Baechtold}, {Barcia}, {Beintema}, {Belitsky}, {Benz}, {Bieber},
  {Boogert}, {Borys}, {Bumble}, {Ca{\"i}s}, {Caris}, {Cerulli-Irelli},
  {Chattopadhyay}, {Cherednichenko}, {Ciechanowicz}, {Coeur-Joly}, {Comito},
  {Cros}, {de Jonge}, {de Lange}, {Delforges}, {Delorme}, {den Boggende},
  {Desbat}, {Diez-Gonz{\'a}lez}, {di Giorgio}, {Dubbeldam}, {Edwards},
  {Eggens}, {Erickson}, {Evers}, {Fich}, {Finn}, {Franke}, {Gaier}, {Gal},
  {Gao}, {Gallego}, {Gauffre}, {Gill}, {Glenz}, {Golstein}, {Goulooze},
  {Gunsing}, {G{\"u}sten}, {Hartogh}, {Hatch}, {Higgins}, {Honingh}, {Huisman},
  {Jackson}, {Jacobs}, {Jacobs}, {Jarchow}, {Javadi}, {Jellema}, {Justen},
  {Karpov}, {Kasemann}, {Kawamura}, {Keizer}, {Kester}, {Klapwijk}, {Klein},
  {Kollberg}, {Kooi}, {Kooiman}, {Kopf}, {Krause}, {Krieg}, {Kramer},
  {Kruizenga}, {Kuhn}, {Laauwen}, {Lai}, {Larsson}, {Leduc}, {Leinz}, {Lin},
  {Liseau}, {Liu}, {Loose}, {L{\'o}pez-Fernandez}, {Lord}, {Luinge}, {Marston},
  {Mart{\'{\i}}n-Pintado}, {Maestrini}, {Maiwald}, {McCoey}, {Mehdi}, {Megej},
  {Melchior}, {Meinsma}, {Merkel}, {Michalska}, {Monstein}, {Moratschke},
  {Morris}, {Muller}, {Murphy}, {Naber}, {Natale}, {Nowosielski}, {Nuzzolo},
  {Olberg}, {Olbrich}, {Orfei}, {Orleanski}, {Ossenkopf}, {Peacock}, {Pearson},
  {Peron}, {Phillip-May}, {Piazzo}, {Planesas}, {Rataj}, {Ravera}, {Risacher},
  {Salez}, {Samoska}, {Saraceno}, {Schieder}, {Schlecht}, {Schl{\"o}der},
  {Schm{\"u}lling}, {Schultz}, {Schuster}, {Siebertz}, {Smit}, {Szczerba},
  {Shipman}, {Steinmetz}, {Stern}, {Stokroos}, {Teipen}, {Teyssier}, {Tils},
  {Trappe}, {van Baaren}, {van Leeuwen}, {van de Stadt}, {Visser}, {Wildeman},
  {Wafelbakker}, {Ward}, {Wesselius}, {Wild}, {Wulff}, {Wunsch}, {Tielens},
  {Zaal}, {Zirath}, {Zmuidzinas}, \& {Zwart}}]{Graauw2010}
{de Graauw}, T., {Helmich}, F.~P., {Phillips}, T.~G., {et~al.} 2010, \aap, 518,
  L6

\bibitem[{{Dobashi}(2011)}]{Dobashi2011}
{Dobashi}, K. 2011, \pasj, 63, S1

\bibitem[{{Dobashi} {et~al.}(1994){Dobashi}, {Bernard}, {Yonekura}, \&
  {Fukui}}]{Dobashi1994}
{Dobashi}, K., {Bernard}, J.-P., {Yonekura}, Y., \& {Fukui}, Y. 1994, \apjs,
  95, 419

\bibitem[{{Dobashi} {et~al.}(2005){Dobashi}, {Uehara}, {Kandori}, {Sakurai},
  {Kaiden}, {Umemoto}, \& {Sato}}]{Dobashi2005}
{Dobashi}, K., {Uehara}, H., {Kandori}, R., {et~al.} 2005, \pasj, 57, S1

\bibitem[{{Ducati}(2002)}]{2002yCat.2237....0D}
{Ducati}, J.~R. 2002, VizieR Online Data Catalog, 2237

\bibitem[{{Dunham} {et~al.}(2011){Dunham}, {Rosolowsky}, {Evans}, {Cyganowski},
  \& {Urquhart}}]{Dunham2011}
{Dunham}, M.~K., {Rosolowsky}, E., {Evans}, II, N.~J., {Cyganowski}, C., \&
  {Urquhart}, J.~S. 2011, \apj, 741, 110

\bibitem[{{ESA}(1997)}]{1997ESASP1200.....E}
{ESA}, ed. 1997, ESA Special Publication, Vol. 1200, {The HIPPARCOS and TYCHO
  catalogues. Astrometric and photometric star catalogues derived from the ESA
  HIPPARCOS Space Astrometry Mission}

\bibitem[{{Federman} {et~al.}(1994){Federman}, {Strom}, {Lambert}, {Cardelli},
  {Smith}, \& {Joseph}}]{Federman1994}
{Federman}, S.~R., {Strom}, C.~J., {Lambert}, D.~L., {et~al.} 1994, \apj, 424,
  772

\bibitem[{{Federman} \& {Willson}(1982)}]{1982ApJ...260..124F}
{Federman}, S.~R., \& {Willson}, R.~F. 1982, \apj, 260, 124

\bibitem[{{Federman} \& {Willson}(1984)}]{1984ApJ...283..626F}
---. 1984, \apj, 283, 626

\bibitem[{{Goldsmith} {et~al.}(2010){Goldsmith}, {Velusamy}, {Li}, \&
  {Langer}}]{2010ApJ...715.1370G}
{Goldsmith}, P.~F., {Velusamy}, T., {Li}, D., \& {Langer}, W.~D. 2010, \apj,
  715, 1370

\bibitem[{{Gredel} {et~al.}(1994){Gredel}, {van Dishoeck}, \&
  {Black}}]{1994A&A...285..300G}
{Gredel}, R., {van Dishoeck}, E.~F., \& {Black}, J.~H. 1994, \aap, 285, 300

\bibitem[{{Gredel} {et~al.}(1992){Gredel}, {van Dishoeck}, {de Vries}, \&
  {Black}}]{1992A&A...257..245G}
{Gredel}, R., {van Dishoeck}, E.~F., {de Vries}, C.~P., \& {Black}, J.~H. 1992,
  \aap, 257, 245

\bibitem[{{Guetter}(1968)}]{1968PASP...80..197G}
{Guetter}, H.~H. 1968, \pasp, 80, 197

\bibitem[{{Hirschauer} {et~al.}(2009){Hirschauer}, {Federman}, {Wallerstein},
  \& {Means}}]{2009ApJ...696.1533H}
{Hirschauer}, A., {Federman}, S.~R., {Wallerstein}, G., \& {Means}, T. 2009,
  \apj, 696, 1533

\bibitem[{{H{\o}g} {et~al.}(2000){H{\o}g}, {Fabricius}, {Makarov}, {Urban},
  {Corbin}, {Wycoff}, {Bastian}, {Schwekendiek}, \&
  {Wicenec}}]{2000A&A...355L..27H}
{H{\o}g}, E., {Fabricius}, C., {Makarov}, V.~V., {et~al.} 2000, \aap, 355, L27

\bibitem[{{Hohle} {et~al.}(2010){Hohle}, {Neuh{\"a}user}, \&
  {Schutz}}]{2010AN....331..349H}
{Hohle}, M.~M., {Neuh{\"a}user}, R., \& {Schutz}, B.~F. 2010, Astronomische
  Nachrichten, 331, 349

\bibitem[{{Houk} \& {Smith-Moore}(1988)}]{1988mcts.book.....H}
{Houk}, N., \& {Smith-Moore}, M. 1988, {Michigan Catalogue of Two-dimensional
  Spectral Types for the HD Stars. Volume 4, Declinations -26.0 deg to -12.0
  deg.}

\bibitem[{{Houk} \& {Swift}(1999)}]{1999MSS...C05....0H}
{Houk}, N., \& {Swift}, C. 1999, in Michigan Spectral Survey, Ann Arbor, Dep.
  Astron., Univ. Michigan, Vol. 5, p. 0 (1999), Vol.~5, 0

\bibitem[{{Ingalls} {et~al.}(2011){Ingalls}, {Bania}, {Boulanger}, {Draine},
  {Falgarone}, \& {Hily-Blant}}]{2011ApJ...743..174I}
{Ingalls}, J.~G., {Bania}, T.~M., {Boulanger}, F., {et~al.} 2011, \apj, 743,
  174

\bibitem[{{Jaschek} \& {Egret}(1982)}]{1982IAUS...98..261J}
{Jaschek}, M., \& {Egret}, D. 1982, in IAU Symposium, Vol.~98, Be Stars, ed.
  M.~{Jaschek} \& H.-G. {Groth}, 261

\bibitem[{{Kim} {et~al.}(2004){Kim}, {Kawamura}, {Yonekura}, \&
  {Fukui}}]{Kim2004}
{Kim}, B.~G., {Kawamura}, A., {Yonekura}, Y., \& {Fukui}, Y. 2004, \pasj, 56,
  313

\bibitem[{{Knapp} \& {Jura}(1976)}]{1976ApJ...209..782K}
{Knapp}, G.~R., \& {Jura}, M. 1976, \apj, 209, 782

\bibitem[{{Lambert} {et~al.}(1990){Lambert}, {Sheffer}, \&
  {Crane}}]{1990ApJ...359L..19L}
{Lambert}, D.~L., {Sheffer}, Y., \& {Crane}, P. 1990, \apjl, 359, L19

\bibitem[{{Langer} {et~al.}(2010){Langer}, {Velusamy}, {Pineda}, {Goldsmith},
  {Li}, \& {Yorke}}]{Langer2010}
{Langer}, W.~D., {Velusamy}, T., {Pineda}, J.~L., {et~al.} 2010, \aap, 521, L17

\bibitem[{{Langer} {et~al.}(2014){Langer}, {Velusamy}, {Pineda}, {Willacy}, \&
  {Goldsmith}}]{2014A&A...561A.122L}
{Langer}, W.~D., {Velusamy}, T., {Pineda}, J.~L., {Willacy}, K., \&
  {Goldsmith}, P.~F. 2014, \aap, 561, A122

\bibitem[{{Liszt}(1979)}]{1979ApJ...233L.147L}
{Liszt}, H.~S. 1979, \apjl, 233, L147

\bibitem[{{Lynds}(1962)}]{Lynds1962}
{Lynds}, B.~T. 1962, \apjs, 7, 1

\bibitem[{{Mattila}(1986)}]{1986A&A...160..157M}
{Mattila}, K. 1986, \aap, 160, 157

\bibitem[{{Morgan} {et~al.}(1955){Morgan}, {Code}, \&
  {Whitford}}]{1955ApJS....2...41M}
{Morgan}, W.~W., {Code}, A.~D., \& {Whitford}, A.~E. 1955, \apjs, 2, 41

\bibitem[{{Pan} {et~al.}(2005){Pan}, {Federman}, {Sheffer}, \&
  {Andersson}}]{Pan2005}
{Pan}, K., {Federman}, S.~R., {Sheffer}, Y., \& {Andersson}, B.-G. 2005, \apj,
  633, 986

\bibitem[{{Pilbratt} {et~al.}(2010){Pilbratt}, {Riedinger}, {Passvogel},
  {Crone}, {Doyle}, {Gageur}, {Heras}, {Jewell}, {Metcalfe}, {Ott}, \&
  {Schmidt}}]{Pilbratt2010}
{Pilbratt}, G.~L., {Riedinger}, J.~R., {Passvogel}, T., {et~al.} 2010, \aap,
  518, L1

\bibitem[{{Pineda} {et~al.}(2013){Pineda}, {Langer}, {Velusamy}, \&
  {Goldsmith}}]{2013A&A...554A.103P}
{Pineda}, J.~L., {Langer}, W.~D., {Velusamy}, T., \& {Goldsmith}, P.~F. 2013,
  \aap, 554, A103

\bibitem[{{Schlingman} {et~al.}(2011){Schlingman}, {Shirley}, {Schenk},
  {Rosolowsky}, {Bally}, {Battersby}, {Dunham}, {Ellsworth-Bowers}, {Evans},
  {Ginsburg}, \& {Stringfellow}}]{Schlingman2011}
{Schlingman}, W.~M., {Shirley}, Y.~L., {Schenk}, D.~E., {et~al.} 2011, \apjs,
  195, 14

\bibitem[{{Sheffer} {et~al.}(2008){Sheffer}, {Rogers}, {Federman}, {Abel},
  {Gredel}, {Lambert}, \& {Shaw}}]{Sheffer2008}
{Sheffer}, Y., {Rogers}, M., {Federman}, S.~R., {et~al.} 2008, \apj, 687, 1075

\bibitem[{{Stil} {et~al.}(2006){Stil}, {Taylor}, {Dickey}, {Kavars}, {Martin},
  {Rothwell}, {Boothroyd}, {Lockman}, \& {McClure-Griffiths}}]{Stil2006}
{Stil}, J.~M., {Taylor}, A.~R., {Dickey}, J.~M., {et~al.} 2006, \aj, 132, 1158

\bibitem[{{Tull} {et~al.}(1995){Tull}, {MacQueen}, {Sneden}, \&
  {Lambert}}]{Tull1995}
{Tull}, R.~G., {MacQueen}, P.~J., {Sneden}, C., \& {Lambert}, D.~L. 1995,
  \pasp, 107, 251

\bibitem[{{Velusamy} {et~al.}(2010){Velusamy}, {Langer}, {Pineda}, {Goldsmith},
  {Li}, \& {Yorke}}]{2010A&A...521L..18V}
{Velusamy}, T., {Langer}, W.~D., {Pineda}, J.~L., {et~al.} 2010, \aap, 521, L18

\bibitem[{{Walborn} \& {Fitzpatrick}(2000)}]{2000PASP..112...50W}
{Walborn}, N.~R., \& {Fitzpatrick}, E.~L. 2000, \pasp, 112, 50

\bibitem[{{Welsh} \& {Sallmen}(2003)}]{2003A&A...408..545W}
{Welsh}, B.~Y., \& {Sallmen}, S. 2003, \aap, 408, 545

\bibitem[{{Welty} \& {Hobbs}(2001)}]{2001ApJS..133..345W}
{Welty}, D.~E., \& {Hobbs}, L.~M. 2001, \apjs, 133, 345

\bibitem[{{Welty} {et~al.}(2003){Welty}, {Hobbs}, \& {Morton}}]{Welty2003}
{Welty}, D.~E., {Hobbs}, L.~M., \& {Morton}, D.~C. 2003, \apjs, 147, 61

\bibitem[{{Wenger} {et~al.}(2000){Wenger}, {Ochsenbein}, {Egret}, {Dubois},
  {Bonnarel}, {Borde}, {Genova}, {Jasniewicz}, {Lalo{\"e}}, {Lesteven}, \&
  {Monier}}]{Wenger2000}
{Wenger}, M., {Ochsenbein}, F., {Egret}, D., {et~al.} 2000, \aaps, 143, 9

\bibitem[{{Willson}(1981)}]{1981ApJ...247..116W}
{Willson}, R.~F. 1981, \apj, 247, 116

\bibitem[{{Wouterloot} \& {Brand}(1989)}]{Wouterloot1989}
{Wouterloot}, J.~G.~A., \& {Brand}, J. 1989, \aaps, 80, 149

\bibitem[{{Wouterloot} {et~al.}(1988){Wouterloot}, {Walmsley}, \&
  {Henkel}}]{Wouterloot1988}
{Wouterloot}, J.~G.~A., {Walmsley}, C.~M., \& {Henkel}, C. 1988, \aap, 203, 367

\bibitem[{{Zhang} {et~al.}(2005){Zhang}, {Zhang}, \& {Li}}]{Zhang2005}
{Zhang}, X.-B., {Zhang}, R.-X., \& {Li}, Z.-P. 2005, \cjaa, 29, 9Z

\end{thebibliography}

\end{document}